\documentclass[10pt, twocolumn]{IEEEtran}

\IEEEoverridecommandlockouts

\usepackage{flushend}
\usepackage{graphicx}
\usepackage{subfigure}
\usepackage{color}
\usepackage{epsfig}
\usepackage{amssymb}
\usepackage{amsmath}
\usepackage{amsthm}
\usepackage{latexsym}

\newcommand{\bPP}[1]{{\mathtt{P}}_{#1}}
\newcommand{\bPr}[1]{{\mathbb{P}}\left(#1\right)}

\newcommand{\bP}[2]{{\mathtt{P}}_{#1}\left({#2}\right)}

\newcommand{\cX}{{\mathcal X}}
\newcommand{\cY}{{\mathcal Y}}
\newcommand{\cE}{{\mathcal E}}
\newcommand{\cF}{{\mathcal F}}
\newcommand{\cW}{{\mathcal W}}

\newcommand{\cU}{{\mathcal U}}
\newcommand{\cK}{{\mathcal K}}
\newcommand{\cV}{{\mathcal V}}

\newcommand{\bF}{\mathbf{F}}
\newcommand{\bx}{\mathbf{x}}
\newcommand{\by}{\mathbf{y}}
\newcommand{\bu}{\mathbf{u}}
\newcommand{\bv}{\mathbf{v}}

\newcommand{\mc}{-\!\!\!\!\circ\!\!\!\!-}
\newcommand{\nn}{\nonumber}

\newcommand{\ep}{\epsilon}

\newtheorem{theorem}{Theorem}

\newtheorem*{corollary}{Corollary}

\newtheorem{lemma}[theorem]{Lemma}

\theoremstyle{remark}
\newtheorem*{remark*}{Remark}
\newtheorem*{remarks*}{Remarks}

\theoremstyle{definition}
\newtheorem{definition}{Definition}

\begin{document}

\title{Common Information and Secret Key Capacity}

\author{Himanshu Tyagi$^\dag$}
\maketitle {\renewcommand{\thefootnote}{}\footnotetext{

\vspace{.02in}\noindent This work was supported by the U.S.
National Science Foundation under Grants CCF0830697 and
CCF1117546.

\noindent$^\dag$Department of Electrical and Computer Engineering,
 and Institute for Systems Research, University of Maryland, College
 Park, MD 20742, USA. Email: tyagi@umd.edu.
 
 A preliminary version of this paper was presented at the IEEE International
 Symposium on Information Theory, St. Petersburg, Russia, July 31 - August 5, 2011.
}}

\begin{abstract}
We study the generation of a secret key of maximum rate by a pair of
terminals observing correlated sources and with the means to communicate
over a noiseless public communication channel. Our main result
establishes a structural equivalence between the generation of a
maximum rate secret key and the generation of a common randomness
that renders the observations of the 
two terminals conditionally
independent. The minimum rate of such common randomness, termed
interactive common information, is related to
Wyner's notion of common information, and serves to characterize
the minimum rate of interactive public communication required to
generate an optimum rate secret key. This characterization yields
a single-letter expression for the aforementioned
communication rate when the number of rounds of interaction are
bounded. An application of our results shows that interaction does
not reduce this rate for binary symmetric
sources. Further, we provide an example for which
interaction does reduce the 
minimum rate of communication. Also, 
certain invariance properties of common information
quantities are established that may be of independent interest.
\end{abstract}

\section{Introduction}
Consider secret key (SK) generation by a pair of terminals that
observe independent and identically distributed (i.i.d.)
repetitions of two discrete, finite-valued random variables (rvs) of
known joint probability mass function. The terminals communicate over a
noiseless public channel of unlimited capacity, interactively in
multiple rounds, to agree upon the value of the key. The key is required
to be (almost) independent of the public communication.
The maximum rate of such an SK, termed the secret key capacity, 
was characterized in \cite{Mau93,
AhlCsi93}.

In the works of Maurer and Ahlswede-Csisz{\'a}r \cite{Mau93,
AhlCsi93}, SK generation of maximum rate entailed both the
terminals recovering the observations of one of the terminals,
using the least rate of communication required to do so. Later, it
was shown by Csisz{\'a}r-Narayan \cite{CsiNar04} that a maximum rate
SK can be generated also by the terminals recovering the observations
of both the terminals. Clearly, the latter scheme requires more
communication than the former. In this paper, we address the
following question, which was raised in
\cite[Section VI]{CsiNar04}: \\
{\it What is the minimum overall rate of interactive communication $R_{SK}$
required to establish
 a maximum rate SK?}

We answer this question by characterizing the form of common
randomness (CR) (i.e., shared bits, see \cite{AhlCsi98}) 
that the terminals must establish
in order to generate a maximum rate SK; 
two examples of such common randomness are the observations of any
one terminal \cite{Mau93,AhlCsi93} and of both terminals \cite{CsiNar04}.
While our main result does
not 
yield 
a single-letter characterization, it nonetheless
reveals a central link between secrecy generation and Wyner's
notion of common information (CI) between two dependent rvs
$X$ and $Y$ 
\cite{Wyn75}. Wyner defined CI as the minimum rate of a function
 of i.i.d. repetitions of 
 two correlated random variables $X$ and $Y$ that facilitated 
a certain distributed source coding task. Alternatively, it can be
defined as the minimum rate of a function of i.i.d. repetitions of
$X$ and $Y$ such that, conditioned on this function, the i.i.d. sequences
are
(almost) independent; this definition, though not stated
explicitly in \cite{Wyn75}, follows from the analysis therein. We
introduce a variant of this notion of CI called the
\emph{interactive CI} where we seek the minimum rate of CR that
renders the mentioned sequences conditionally independent.
Clearly, interactive CI cannot be smaller than Wyner's CI, and can exceed it.
Our main contribution is to show a one-to-one correspondence
between such CR and the CR established for generating an optimum
rate SK. This correspondence is used to characterize the
minimum rate of communication $R_{SK}$ required for generating 
a maximum
rate SK. In fact, it is shown that 
 $R_{SK}$ 
is simply
interactive CI minus the secret key capacity.

When the number of rounds of interaction are bounded, this
characterization yields a single-letter expression for 
$R_{SK}$.
Using this expression 
we show that an interactive communication scheme can have less rate than
a noninteractive one, in general. However, interaction offers no advantage 
for binary symmetric sources.
This expression also illustrates the role of sufficient statistics in SK
generation. We further dwell on this relationship and show that many
CI quantities of interest remain unchanged if the sources are
replaced by their corresponding sufficient statistics 
(with respect to each other). Interestingly, the effect of
substitution by sufficient statistics has been studied in the 
context of the 
rate-distortion problem for a remote source 
in \cite[Lemma 2]{EswGas05}, 
and recently, for the lossy and lossless distributed source coding problems
in \cite{XuChen12}. Here, in effect, we study this substitution for the
distributed source coding problems underlying the CI quantities.

The basic notions of CR and SK are explained in the next section. The
definition of interactive CI and the heuristics underlying our approach
are given in Section \ref{s_iCI}. Our main results are provided in
Section \ref{s_res}, followed by illustrative examples in the subsequent
section. Section \ref{s_suff} explores the connection between 
sufficient statistics and common information quantities. A discussion
of our results and possible extensions is given in the final section.

\noindent{\it Notation.} The rvs $X$ and $Y$ take values in finite
sets $\cX$ and $\cY$, respectively. Let
$X^n=(X_1,...,X_n)$ and $Y^n=(Y_1,...,Y_n)$ denote $n$ i.i.d.
repetitions of $X$ and $Y$, respectively. For a collection of rvs
$U_1,..., U_r$, for $i\leq j$ let $U_i^j$ denote $U_i,
U_{i+1}..., U_j$; 
when $i=1$, we use $U^j = U_1, ..., U_j$.
For rvs $U, V$, and $0< \ep< 1$, we say $U$ is
$\ep$-recoverable from $V$ if there is a function $g$ of $V$ such
that
\begin{align}\nn
\bPr{U = g(V)} \geq 1- \ep.
\end{align}
Denote the cardinality of the range space of a mapping $f$ by
$\|f\|$, and similarly, with a slight abuse of notation, the (fixed) 
range space
of a random mapping $\mathbf{F}$ by $\|\mathbf{F}\|$.

\section{Interactive Communication, Common Randomness and Secret Keys}\label{s_def}
Terminals $\cX$ and $\cY$ (with a slight abuse of notation)
communicate interactively,
with, say, terminal $\cX$
transmitting first. Each terminal then
communicates alternately for $r$ rounds. Specifically, an
\emph{$r$-interactive communication} ${\bf f} = \left(f_1, f_2,
..., f_r\right)$ is a sequence of finite-valued mappings
 with
\begin{align}\nn
&f_{2i+1} : \cX^n \times \cF^{2i} \rightarrow \cF_{2i+1},\quad 0
\leq i \leq \lfloor(r-1)/2 \rfloor,
\\\nn &f_{2i} : \cY^n \times \cF^{2i-1} \rightarrow
\cF_{2i},\quad 1 \leq i \leq \lfloor r/2 \rfloor,
\end{align}
where $\left\{\cF_i\right\}_{i = 1}^r$ are finite sets and $\cF_0
= \emptyset$. This set-up subsumes protocols where 
terminal $\cY$
initiates the communication upon choosing $f_1 =$ constant. Let
${\bf F} = \mathbf{f}\left(X^n, Y^n\right)$ 
describe collectively 
the corresponding rv. The rate of this communication is given by
$$\frac{1}{n}\log\|\mathbf{F}\|.$$
We assume that the communication from each
terminal is a (deterministic) function of its knowledge.
In particular, randomization is not allowed. This is 
 not a limiting assumption; see Section \ref{s_discA}.

\begin{definition}
Given interactive communication $\mathbf{F}$ as above, a function
$L$ of
$(X^n, Y^n)$ is \emph{$\ep$-common randomness} ($\ep$-CR)
recoverable from\footnote{The rv $L$ 
is $\ep$-recoverable 
from $(X^n, \mathbf{F})$ or $(Y^n, \mathbf{F})$ but not necessarily from
$\mathbf{F}$ alone. 
The deliberate misuse of the terminology ``recoverable from $\bF$" 
simplifies presentation.
} $\mathbf{F}$
if there exist mappings $L_1 =
L^{(n)}_1(X^n, \mathbf{F})$ and $L_2 = L^{(n)}_2(Y^n, \mathbf{F})$
such that
\begin{align}\nonumber
\bPr{L = L_1 = L_2} \geq 1- \ep.
\end{align}
\end{definition}

\begin{definition}
A function $K$ of $(X^n, Y^n)$, 
with values in a set $\cK$, 
forms an \emph{$\ep$-secret key for $X$ and $Y$}
($\ep$-SK) if $K$ is $\ep$-CR recoverable from $X^n$ or $Y^n$ and (interactive public
communication) $\mathbf{F}$, and
\begin{align}\label{e:weak_secrecy}
\frac{1}{n}I(K \wedge \mathbf{F}) \leq \ep.
\end{align}
For convenience, simplistically, the $\ep$-SK $K$ is said to be recoverable from $\mathbf{F}$.
A rate $R>0$ is an achievable SK rate if for every $0<
\epsilon <1$ there exists, for some\footnote{
Our results hold even if the phrase ``for some $n\geq 1$" is
replaced by ``for all $n$ sufficiently large;" the former has been chosen
here for convenience.
} $n\geq 1$, an $\ep$-SK $K =
K^{(n)}$ with $(1/n)H(K) \geq R-\ep$. The supremum of all
achievable SK rates is denoted by $C$, and is called the SK
capacity.
\end{definition}

The following result\footnote{ It is shown in \cite{Mau94, Csi96} that SK capacity remains unchanged 
even if the notion of ``weak secrecy" of $K$ in (\ref{e:weak_secrecy}) 
is tightened to ``strong secrecy" by omitting the normalization 
with respect to $n$, and an additional uniformity
constraint $H(K) \geq \log|\cK| - \ep$ is imposed. } is well known.
\begin{theorem}\label{t_CSK}\emph{\cite{Mau93, AhlCsi93}}
The SK capacity for $X$ and $Y$ is given by
\begin{align}\label{e:SK_capacity}
C = I (X\wedge Y).
\end{align}
\end{theorem}
\section{Relation between Secret Key and Wyner's Common Information}\label{s_iCI}
We interpret Wyner's CI for a pair of rvs $(X, Y)$ as
the minimum rate of a function of their i.i.d. repetitions
$\left(X^n, Y^n\right)$ that renders $X^n$ and $Y^n$ conditionally
independent. Formally, 

\begin{definition}
$R\geq 0$ is an achievable CI rate
if for every $0< \epsilon <1$ there exists an $n\geq 1$ and
a (finite-valued) rv $L = L\left(X^n, Y^n\right)$ of rate
$(1/n)H(L) \leq R + \ep$ that satisfies the property:
\begin{align}\label{e_CIcond}
\frac{1}{n}I\left(X^n \wedge Y^n \mid L\right) \leq \ep.
\end{align}
\end{definition}

Obvious examples of such an rv $L$ are $L=\left(X^n, Y^n\right)$ or $X^n$
or $Y^n$.
The infimum of all achievable
CI rates, denoted $CI_W(X\wedge Y)$, is called the CI of $X$ and
$Y$. This definition of CI, though not stated explicitly in
\cite{Wyn75}, follows from the analysis therein. The following
theorem characterizes $CI_W(X\wedge Y)$.
\begin{theorem}\label{t_CI}\emph{\cite{Wyn75}} The CI of the rvs $X, Y$
is
\begin{align}\label{e_CIchar}
CI_W(X \wedge Y) = \min_{W}I(X,Y\wedge W),
\end{align}
where the rv $W$ takes values in a (finite) set $\cW$ with $|\cW|
\leq |\cX||\cY|$ and satisfies the Markov condition $X \mc W \mc
Y$.
\end{theorem}
\noindent The direct part follows from \cite[equation
(5.12)]{Wyn75}. The proof of the converse is straightforward. Further,
it is a simple exercise to infer from (\ref{e_CIchar}) that 
$CI_W(X\wedge Y) \geq I(X\wedge Y)$.

\begin{definition}
An achievable \emph{$r$-interactive} CI rate is defined in a manner
analogous to the achievable CI rate, 
but with the restriction that
the rvs $L$
in (\ref{e_CIcond}) be $\ep$-CR, i.e., $L = (J, \mathbf{F})$,
where $\mathbf{F}$ is an $r$-interactive communication and $J$ is
$\ep$-recoverable from $\mathbf{F}$. The infimum of all
achievable $r$-interactive CI rates, denoted $CI_i^r(X;Y)$,
is called the $r$-interactive CI of the rvs $X$ and $Y$. 
By definition, the nonnegative sequence $\left\{CI_i^r(X; Y)\right\}_{r=1}^\infty$ is
nonincreasing in $r$ and is bounded below by $CI_W(X\wedge Y)$. Define
\begin{align}\nn
CI_i(X \wedge Y) = \lim_{r\rightarrow \infty} CI_i^r(X; Y).
\end{align}
\end{definition}
\vskip2ex

Then $CI_i(X \wedge Y) \geq CI_W(X \wedge Y) \geq 0$. Note that
$CI_i^r(X;Y)$ may not be symmetric in $X$ and $Y$ since 
the communication is initiated at terminal $\cX$.
However, since 
$$CI_i^{r+1}(X;Y) \leq CI_i^r(Y;X) \leq CI_i^{r-1}(X;Y), $$
clearly,
\begin{align}\nn
CI_i(X \wedge Y) &= \lim_{r\rightarrow \infty} CI_i^r(X; Y) 
\\\nn &= \lim_{r\rightarrow \infty} CI_i^r(Y; X) 
\\\label{e:CIi_symmetry} &= CI_i(Y \wedge X).
\end{align}
Further, for all $0< \epsilon <1$, $J = X^n$ is $\ep$-recoverable from $Y^n$
and a communication (of a Slepian-Wolf codeword) $F = F\left(X^n\right)$, and $L=(J, F)$
 satisfies (\ref{e_CIcond}). Hence, $CI_i(X \wedge Y) \leq H(X)$;
 similarly, $CI_i(X \wedge Y) \leq H(Y)$.  To
 summarize, we have
\begin{align}\label{e_CIineq}
0\leq CI_W(X \wedge Y) \leq CI_i(X \wedge Y) \leq \min \{H(X),
H(Y)\},
 \end{align}
where the first and the last inequalities can be strict. In
Section \ref{s_BS} we show that the second inequality is strict
for binary symmetric rvs $X, Y$.

The $r$-interactive CI plays a pivotal role in optimum rate SK
generation. 
Loosely speaking, 
our main result asserts the following. \emph{A CR that satisfies
(\ref{e_CIcond}) can be used to generate an optimum rate SK and
conversely, an optimum rate SK yields a CR satisfying
(\ref{e_CIcond}). In fact, such a CR of rate $R$ can be recovered
from an interactive communication of rate $R-C$, where $C$ is the
SK capacity for $X$ and $Y$. Therefore, to find the minimum rate of interactive
communication needed to generate an optimum rate SK, it is
sufficient to characterize $CI_i(X\wedge Y)$.}

\section{Main Results}\label{s_res}

\begin{definition}\label{d:comm_rate}
A rate $R^\prime \geq 0$ is an achievable $r$-interactive
communication rate for $CI_i^r$ if, for all $0< \epsilon <1$,
there exists, for some $n \geq 1$, an $r$-interactive
communication $\mathbf{F}$ of rate $(1/n)\log \|\mathbf{F}\|\leq R^\prime
+ \ep$, and an $\ep$-CR $J$ recoverable from $\mathbf{F}$, with
$L = (J, \mathbf{F})$ satisfying (\ref{e_CIcond}). Let $R_{CI}^r$
denote the infimum of all achievable $r$-interactive communication
rates for $CI_i^r$. Similarly, $R^{\prime\prime}\geq 0$ is an achievable
$r$-interactive communication rate for SK capacity if, for all $0<
\epsilon <1$, there exists, for some $n \geq 1$, an
$r$-interactive communication $\mathbf{F}$ of rate $(1/n)\log
\|\mathbf{F}\|\leq R^{\prime\prime} + \ep$, and an $\ep$-SK $K$, recoverable from
$\mathbf{F}$, of rate $(1/n)H(K)\geq I(X \wedge Y) - \ep$;
 $R_{SK}^r$ denotes the infimum  of all achievable $r$-interactive communication rates for SK capacity.
Note that by their definitions, both $R_{CI}^r$ and $R^r_{SK}$ are
nonincreasing with increasing $r$, and are bounded below by zero.
Define $$\displaystyle R_{CI} = \lim_{r\rightarrow\infty}R_{CI}^r, \quad R_{SK} = \lim_{r\rightarrow\infty}R_{SK}^r.$$
\end{definition}
\vskip2ex

Although $R_{CI}^r(X;Y)$ and $R_{SK}^r(X;Y)$ are not equal to $R_{CI}^r(Y;X)$ and $R_{SK}^r(Y;X)$, respectively,
the quantities $R_{CI}$ and $R_{SK}$ are symmetric in $X$ and $Y$ using 
an argument similar to the one leading to (\ref{e:CIi_symmetry}).

\begin{theorem}\label{t_main}
For every $r\geq 1$,
\begin{align}\label{e_main2}
R_{SK}^r = R_{CI}^r = CI_i^r(X; Y) - I(X \wedge Y).
\end{align}
\end{theorem}
\begin{corollary}\label{c_main}
It holds that
\begin{align}\label{e_main}
R_{SK} = R_{CI} = CI_i(X \wedge Y) - I(X \wedge Y).
\end{align}
\end{corollary}
\begin{remark*}
The relation (\ref{e_main}) can be interpreted as follows. Any CR $J$
recoverable from (interactive communication) $\mathbf{F}$, with $L
= (J, \mathbf{F})$ satisfying (\ref{e_CIcond}), can be decomposed
into two mutually independent parts: An SK $K$ of maximum rate and
the interactive communication $\mathbf{F}$. It follows upon
rewriting (\ref{e_main}) as $CI_i(X \wedge Y) =  I(X \wedge Y) +
R_{CI} $ that the communication $\mathbf{F}$ is (approximately) of
rate $R_{CI}$. Furthermore, $R_{CI}$ is same as $R_{SK}$.
\end{remark*}

A computable characterization of the operational term $CI_i(X
\wedge Y)$ is not known. However, the next result gives a
single-letter characterization of $CI_i^r(X; Y)$.
\begin{theorem}\label{t_char}
Given rvs $X,Y$ and $r \geq 1$, we have
\begin{align}\label{e_iCIr}
CI_i^r(X; Y) = \min_{U_1, ..., U_r} I(X,Y \wedge U_1, ..., U_r),
\end{align}
where the minimum is taken over rvs $U_1, ..., U_r$ taking values
in finite sets $\cU_1, ...,\cU_r$, respectively, that satisfy the
following conditions
\begin{align}\nn
(P1)&\,\,\, U_{2i+1} \mc X, U^{2i} \mc Y, \quad 0 \leq i \leq
\lfloor(r-1)/2 \rfloor,
\\\nn& U_{2i} \mc Y, U^{2i-1}
\mc X,\quad 1 \leq i \leq \lfloor r/2 \rfloor,
\\\nn (P2)&\,\,\, X
\mc  U^{r} \mc Y,
\\\nn
(P3)&\,\,\, |\cU_{2i+1}|\leq |\cX|\prod_{j=1}^{2i}|\cU_j| +
1,\quad 0 \leq i \leq \lfloor(r-1)/2 \rfloor, \\\nn&
|\cU_{2i}|\leq |\cY|\prod_{j=1}^{2i-1}|\cU_j| + 1, \quad 1
\leq i \leq \lfloor r/2 \rfloor,
\end{align}
with $\cU_0 = \emptyset$ and $U_0 =$ constant.
\end{theorem}
\begin{remark*} Note that (\ref{e_iCIr}) has the same form as
the expression for $CI_W(X\wedge Y)$ in (\ref{e_CIchar}) with $W$ replaced by $(U_1, ..., U_r)$ satisfying
the conditions above.
\end{remark*}
Before presenting the proof of our main Theorems \ref{t_main}
and \ref{t_char}, we give some technical results that will
constitute central tools for the proofs.

\begin{lemma}\label{l:Comm_int}
For an interactive communication
$\mathbf{F}$ it holds that
\begin{align}\label{ec1d}
H(\mathbf{F} \mid X^n) + H(\mathbf{F} \mid Y^n) \leq
H(\mathbf{F}).
\end{align}
\end{lemma}

\begin{lemma}\label{l:int_comm_rate}
For an $r$-interactive communication
$\mathbf{F}$, define
\begin{align}\nn
\mathbf{F}_i = \mathbf{F}\left(X_{n(i-1)+1}^{ni}, Y_{n(i-1)+1}^{ni}\right),\quad 1\leq i\leq k.
\end{align}
Then, for all $k \geq k_0(n,\ep, |\cX|, |\cY|)$ there exists an $r$-interactive communication 
$\mathbf{F}^\prime=\mathbf{F}^\prime\left(X^{nk}, Y^{nk}\right)$ of rate 
\begin{align}\label{e:int_comm_rate}
\frac{1}{nk}\log\|\mathbf{F}^\prime\|\leq \frac{1}{n}\left[H\left(\mathbf{F}|X^n\right) + H\left(\mathbf{F}|Y^n\right)\right] + \ep,
\end{align}
such that $\mathbf{F}^k$ is an $\ep$-CR recoverable from $\mathbf{F}^\prime$.
\end{lemma}
\begin{remark*}
Lemma \ref{l:int_comm_rate} says that, in essence, for an optimum rate communication $\mathbf{F}$,
$$\frac{1}{n}\log\|\mathbf{F}\| \approx \frac{1}{n}\left[H\left(\mathbf{F}|X^n\right) + H\left(\mathbf{F}|Y^n\right)\right].$$
\end{remark*}
\begin{lemma}({\it A General Decomposition})\label{l:decomp}
For a CR $J$ recoverable from an interactive communication
$\mathbf{F}$ we have
\begin{align}\nn
&nI(X\wedge Y) 
\\\nn &= I\left(X^n\wedge Y^n\mid J, \mathbf{F}\right) +
H(J,\mathbf{F}) - H\left(\mathbf{F}\mid X^n\right) \\\label{e_eq} &\qquad
-H\left(\mathbf{F}\mid Y^n\right) - H\left(J\mid X^n, \mathbf{F}\right)  - H\left(J\mid Y^n,
\mathbf{F}\right).
\end{align}
\end{lemma}
\noindent Lemma \ref{l:Comm_int} is a special case of 
\cite[Lemma B.1]{CsiNar08} (also, see \cite{MadTet10}). 
The proofs of Lemma \ref{l:int_comm_rate} and Lemma \ref{l:decomp} 
are given in the Appendix. 

Note that a simplification of (\ref{e_eq}) gives
\begin{align}\nn
I(X\wedge Y) &\leq \frac{1}{n}\bigg[I\left(X^n\wedge Y^n\mid J, \mathbf{F}\right) +
\\\label{e_eq1b} &\qquad
H(J,\mathbf{F}) - H\left(\mathbf{F}\mid X^n\right)  -
H\left(\mathbf{F}\mid Y^n\right) \bigg].
\end{align}
If $J$ is an $\ep$-CR recoverable from $\mathbf{F}$, Fano's
inequality implies
\begin{align}\nn
\frac{1}{n}\big[H(J\mid X^n,\mathbf{F}) + H(J\mid Y^n,
\mathbf{F})\big] &\leq 2\ep\log|\cX||\cY| + 2h(\ep) 
\\\label{e_Fano} &= \delta(\ep),\text{ say,}
\end{align}
where $h(\ep) = -\ep\log \ep -(1-\ep)\log(1-\ep)$, and
$\delta(\ep) \rightarrow 0$ as $\ep \rightarrow 0$. Combining
(\ref{e_eq}) and (\ref{e_Fano}) we get
\begin{align}\nn
I(X\wedge Y) &\geq \frac{1}{n}\bigg[I\left(X^n\wedge Y^n\mid J, \mathbf{F}\right) +
H(J,\mathbf{F}) 
\\\label{e_eq1} &\qquad- H\left(\mathbf{F}\mid X^n\right)  -
H\left(\mathbf{F}\mid Y^n\right) \bigg] - \delta(\ep),
\end{align}
and further, by (\ref{ec1d}),
\begin{align}\nn
I(X\wedge Y) &\geq \frac{1}{n}\left[I\left(X^n\wedge Y^n\mid J, \mathbf{F}\right) +
H(J, \mathbf{F}) - H(\mathbf{F})\right] 
\\\label{e_eq2}&\hspace*{5cm}- \delta(\ep).
\end{align}

\subsection{Proof of Theorem \ref{t_main}}
In this section we give a proof for (\ref{e_main2}). The proof of
(\ref{e_main}) then follows upon taking limit $r \rightarrow
\infty$ on both sides of (\ref{e_main2}). The proof of  
(\ref{e_main2}) follows from claims 1-3 below. In particular,
the proofs of claims 1-3 establish a structural equivalence 
between a maximum rate SK and an SK of rate 
$\approx \frac{1}{n}H(J \mid \mathbf{F})$ extracted 
from a CR $J$ recoverable from $\mathbf{F}$ such that
$L=(J, \mathbf{F})$ satisfies (\ref{e_CIcond}).

\noindent \textbf{Claim 1:} $R_{CI}^r \geq CI_i^r(X;Y) - I(X
\wedge Y)$.

\noindent \emph{Proof.} By the definition of $R_{CI}^r$, for every
$0< \ep < 1$ there exists, for some $n\geq 1$, an $r$-interactive
communication $\mathbf{F}$ of rate
\begin{align}\label{ec1a}
\frac{1}{n}\log\|\mathbf{F}\| \leq R_{CI}^r + \ep,
\end{align}
and $J$, an $\ep$-CR recoverable from $\mathbf{F}$, such that $L =
(J, \mathbf{F})$ satisfies (\ref{e_CIcond}).
It follows upon rearranging the terms in (\ref{e_eq2}) that
\begin{align}\nn
\frac{1}{n}H(J, \mathbf{F})&\leq I(X\wedge Y) + \frac{1}{n}H(\mathbf{F}) + \delta(\ep),
\end{align}
which with (\ref{ec1a}) gives
\begin{align}\label{ec1b}
\frac{1}{n}H(J, \mathbf{F}) \leq I(X \wedge Y) + R_{CI}^r + \ep +
\delta(\ep).
\end{align}
Since $(J, \mathbf{F})$ satisfies
$$\frac{1}{n}I\left(X^n \wedge Y^n \mid
J, \mathbf{F}\right) \leq \ep \leq \ep+\delta(\ep),
$$
the inequality (\ref{ec1b}), along with the fact that $(\ep +\delta(\ep))\rightarrow 0$ as
 $\ep \rightarrow 0$, implies that $ I(X \wedge Y) + R_{CI}^r$ is an achievable $r$-interactive
 CI rate; hence, $CI_i^r(X;Y) \leq I(X \wedge Y) + R_{CI}^r$.
\vspace{0.3cm}

\noindent \textbf{Claim 2:} $R_{SK}^r \geq R_{CI}^r$.

\noindent \emph{Proof.} Using the definition of $R_{SK}^r$, for
$0< \ep < 1$ there exists, for some $n \geq 1$, an $r$-interactive
communication $\mathbf{F}$ of rate $\frac{1}{n}\log\|\mathbf{F}\|
\leq R_{SK}^r + \ep$, and an $\ep$-SK $K$ recoverable from
$\mathbf{F}$ of rate
\begin{align}\label{ec1e}
\frac{1}{n}H(K)\geq I(X\wedge Y) - \ep.
\end{align}
By choosing $J=K$ in (\ref{e_eq2}) and rearranging
the terms we get,
$$\frac{1}{n} I\left(X^n\wedge Y^n\mid K, \mathbf{F}\right) \leq I(X\wedge Y) - \frac{1}{n} H(K\mid \mathbf{F}) +\delta(\ep).$$
Next, from $(1/n)I(K \wedge \mathbf{F}) < \ep$, we have
\begin{align}\nn
\frac{1}{n} I\left(X^n\wedge Y^n\mid K, \mathbf{F}\right) &\leq
I(X\wedge Y) - \frac{1}{n} H(K) + \ep +\delta(\ep)
\\\nn &\leq 2\ep +\delta(\ep),
\end{align}
where the last inequality follows from (\ref{ec1e}). Since $(2\ep
+\delta(\ep)) \rightarrow 0$ as $\ep \rightarrow 0$, $R_{SK}^r$ is
an achievable $r$-interactive communication rate for $CI_i^r$, and
thus, $R_{SK}^r \geq R_{CI}^r$. \vspace{0.3cm}

{\bf Claim 3:} $R_{SK}^r \leq CI_i^r(X;Y) - I(X \wedge Y)  $.
\vspace{0.1cm}

{\it Proof.} For $0< \ep < 1$,  let $J$ be an $\ep$-CR recoverable
from an $r$-interactive communication $\mathbf{F}$, with
\begin{align}
\frac{1}{n}H(J, \mathbf{F}) \leq CI_i^r(X;Y) + \ep,
\end{align}
such that $L = (J, \mathbf{F})$ satisfies (\ref{e_CIcond}),
and so, by (\ref{e_eq1b}),
\begin{align}\nn
&\frac{1}{n}\left[H(\mathbf{F}\mid X^n) +H(\mathbf{F}\mid Y^n)
\right]
\\\nn &\leq \frac{1}{n}H(J, \mathbf{F}) - I(X\wedge Y) +
\epsilon
\\\label{ec3b} &\leq CI_i^r(X;Y) -  I(X\wedge Y) + 2\epsilon.
\end{align}
To prove the assertion in claim 3, we show that for some $N \geq 1$ there
exists $\Delta(\ep)$-SK $K = K(X^N, Y^N)$ of rate 
$$\frac{1}{n}\log\|K\| \geq I(X\wedge Y) - \Delta(\ep)$$
recoverable from an $r$-interactive communication
$\mathbf{F}^{\prime\prime}= \mathbf{F}^{\prime\prime}(X^N, Y^N)$ of rate
\begin{align}\label{ec3c}
\frac{1}{N}\log\|\mathbf{F}^{\prime\prime}\| \leq
\frac{1}{n}\left[H(\mathbf{F}\mid X^n) +H(\mathbf{F}\mid Y^n)
\right] +\Delta(\ep) - 2\ep,
\end{align}
where 
$\Delta(\ep)\rightarrow 0$ as $\ep
\rightarrow 0$. Then (\ref{ec3c}), along with (\ref{ec3b}),
would yield
\begin{align}\label{ec3e}
\frac{1}{N}\log\|\mathbf{F}^{\prime\prime}\| \leq CI_i^r(X;Y) - I(X\wedge
Y) + \Delta(\ep),
\end{align}
so that $CI_i^r(X;Y) -  I(X\wedge Y)$ is an achievable
$r$-interactive communication rate for SK capacity, 
thereby establishing the claim.

It remains to find $K$ and $\mathbf{F}^{\prime\prime}$ as
above. To that end, let $J$ be recovered as $J_1 = J_1(X^n, \mathbf{F})$ and $J_2 =
J_2(Y^n, \mathbf{F})$ by terminals
$\cX$ and
$\cY$, respectively, i.e., 
\begin{align}\nn
\bPr{J = J_1 = J_2} \geq 1-\ep.
\end{align}
Further, for $k\geq 1$, let
\begin{align}\nn
 J_{1i} &= J_1\left(X_{n(i-1)+1}^{ni}, \mathbf{F}_i\right), 
 \\\nn
J_{2i} &= J_2\left(Y_{n(i-1)+1}^{ni}, \mathbf{F}_i\right), \quad 1\leq i\leq k,
\end{align}
where $\mathbf{F}_i = \mathbf{F}\left(X_{n(i-1)+1}^{ni}, Y_{n(i-1)+1}^{ni}\right)$.
For odd $r$, we find an $r$-interactive communication
$\mathbf{F}^{\prime\prime}$ such that $\left(J_1^k, \mathbf{F}^k\right)$
is a $\ep$-CR recoverable from $\mathbf{F}^{\prime\prime}$,
for all $k$ sufficiently large; the
the SK $K$ will be chosen to be a function of 
$\left(J_1^k, \mathbf{F}^k\right)$ of appropriate rate. 
The proof for even $r$ is similar and is obtained by interchanging the
roles of $J_1$ and $J_2$.
In particular, by
Lemma \ref{l:int_comm_rate}, for all $k$ sufficiently large there exists 
an $r$-interactive communication $\mathbf{F}^\prime$ such that 
$\mathbf{F}^k$ is $\ep$-CR recoverable from $\mathbf{F}^\prime$
of rate given by (\ref{e:int_comm_rate}).
Next, from Fano's
inequality 
\begin{align}\label{ec3f}
\frac{1}{n}\max\{H(J\mid J_1); H(J_1\mid J_2)\} \leq \ep\log
|\cX||\cY| + h(\ep).
\end{align}
By the Slepian-Wolf theorem \cite{SleWol73} there exists a mapping
 $f$ of $J_1^k$ of rate
\begin{align}\label{ec3i}
\frac{1}{k}\log\|f\|\leq H(J_1\mid J_2) +n\ep,
\end{align}
such that
\begin{align}\label{ec3j}
J_1^k \text{ is $\ep$-recoverable from}
\left(f\left(J_1^k\right), J_2^k\right),
\end{align}
for all $k$ sufficiently large. It follows from (\ref{ec3f}), (\ref{ec3i}) that
\begin{align}\label{e:SW_rate}
\frac{1}{nk}\log\|f\|\leq \ep + \ep\log
|\cX||\cY| + h(\ep).
\end{align}
For $N=nk$, we define the $r$-interactive communication 
$\mathbf{F}^{\prime\prime} = \mathbf{F}^{\prime\prime}\left(X^N, Y^N\right)$
as 
\begin{align}\nn
F^{\prime\prime}_i &= F^\prime_i, \quad 1\leq i \leq r-1,\\\nn
F^{\prime\prime}_k &= F^\prime_r, f(J_1^k), \quad i=r,
\end{align}
Thus, $\left(J_1^k,\mathbf{F}^k\right)$ is $2\ep$-CR recoverable from 
$\mathbf{F}^{\prime\prime}$, where, by (\ref{e:int_comm_rate}) and (\ref{e:SW_rate}),
the rate of communication 
$\mathbf{F}^{\prime\prime}$ is bounded by
\begin{align}\nn
&\frac{1}{nk}\log\|\mathbf{F}^{\prime\prime}\|
\\\label{e:F''_rate} &\leq \frac{1}{n}\left[H\left(\mathbf{F}|X^n\right) + H\left(\mathbf{F}|Y^n\right)\right]
 + 2\ep+\ep\log
|\cX||\cY| + h(\ep).
\end{align}

Finally, to construct the SK $K = K\left(J_1^k, \mathbf{F}^k\right)$, using the corollary of Balanced
Coloring Lemma in \cite[Lemma B.3]{CsiNar04}, with
\begin{align}\nn
U = (J_1,\mathbf{F}), \,\, V = \phi,\,\,n=k,\,\, g =
\mathbf{F}^\prime,
\end{align}
we get from (\ref{e:F''_rate}) that there exists a function $K$ of
$J_1^k, \mathbf{F}^k$ such that
\begin{align}\nonumber
&\frac{1}{k}\log \|K\|
\\\nn &\geq 
H(U) - \frac{1}{k}\log\|\mathbf{F}^{\prime\prime}\|
\\\nn  &\geq
H(J_1, \mathbf{F}) - H(\mathbf{F}\mid
X^n) - H(\mathbf{F}\mid Y^n)
\\\label{ec4c} &\hspace*{2cm}- n(2\ep +\ep\log|\cX||\cY| +
h(\ep)),
\end{align}
and
\begin{align}\nn
I(K \wedge \mathbf{F}^\prime)\leq \exp(-ck),
\end{align}
where $c>0$, for all sufficiently large $k$. 
We get from
(\ref{ec4c}) and (\ref{e_eq1b}) that the rate of $K$
is bounded below as follows:
\begin{align}\nn
\frac{1}{nk}\log\|K\|&\geq I(X\wedge Y) -
\frac{1}{n}I\left(X^n \wedge Y^n \mid J_1, \mathbf{F}\right)
\\\label{e:SK_rate}&\qquad -2\ep
-\ep\log|\cX||\cY| - h(\ep).
\end{align}
Observe that
\begin{align}\nn
I(X^n \wedge Y^n \mid J, \mathbf{F}) &= I(J_1, X^n \wedge Y^n \mid
J, \mathbf{F}) \\\nn &\geq I(X^n \wedge Y^n \mid J, J_1,
\mathbf{F})\\\nn &\geq I(X^n \wedge Y^n \mid J_1, \mathbf{F}) -
H(J\mid J_1),
\end{align}
which along with (\ref{ec3f}), and the fact that $L =
(J,\mathbf{F})$ satisfies (\ref{e_CIcond}), yields
\begin{align}\label{ec3ff}
\frac{1}{n}I(X^n \wedge Y^n \mid J_1, \mathbf{F}) \leq \epsilon +
\ep\log |\cX||\cY| + h(\ep).
\end{align}
Upon combining (\ref{e:SK_rate}) and (\ref{ec3ff}) we get,
\begin{align}\nn
\frac{1}{nk}\log\|K\|\geq I(X\wedge Y) -3\ep
-2\ep\log|\cX||\cY| - 2h(\ep).
\end{align}
Thus, for $\Delta(\ep) = 4\ep
+2\ep\log|\cX||\cY| + 2h(\ep)$
$K$ is a $\Delta(\ep)$-SK of rate $(1/nk)\log
\|K\| \geq I(X \wedge Y) -\Delta(\ep)$, recoverable from
$r$-interactive communication $\mathbf{F}^{\prime\prime}$, 
which with (\ref{e:F''_rate}), completes the proof.\qed

\subsection{Proof of Theorem \ref{t_char}}
\emph{Achievability.} Consider rvs $U_1, ..., U_r$ satisfying
conditions (P1)-(P3) in the statement of Theorem \ref{t_char}. 
It suffices to show for
every $0< \ep < 1$, for some $n \geq 1$, there exists an
$r$-interactive communication $\mathbf{F}$, and $\ep$-CR $J$
recoverable from $\mathbf{F}$, such that
\begin{align}\label{el1}
I(X, Y \wedge U^r) - \ep\leq \frac{1}{n}H(J, \mathbf{F}) \leq I(X,
Y \wedge U^r) + \ep,
\end{align}
and
\begin{align}\label{el2}
\frac{1}{n}H(\mathbf{F}) \leq I(X, Y \wedge U^r) - I(X\wedge Y) +
\ep,
\end{align}
since from (\ref{e_eq2}), (\ref{el1}) and (\ref{el2}), we have
\begin{align}\nn
&\frac{1}{n}I\left(X^n \wedge Y^n \mid J, \mathbf{F}\right) 
\\\nn &\leq
\frac{1}{n}H(\mathbf{F}) -\frac{1}{n}H(J, \mathbf{F}) + I(X\wedge
Y) + \delta(\ep)\\\nn &\leq  2\ep+\delta(\ep).
\end{align}
We show below that
\begin{align}
\\\nn &I(X, Y \wedge U^r) - I(X\wedge Y)
\\\nn &= \sum_{i=0}^{\lfloor
(r-1)/2\rfloor} I(X\wedge U_{2i+1}\mid Y, U^{2i})
\\\label{el5} &\qquad +\sum_{i=1}^{\lfloor r/2\rfloor} I(Y\wedge U_{2i}\mid X,
U^{2i-1}).
\end{align}
Thus, the proof will be completed upon showing that there
exists an $\ep$-CR $J$, recoverable from $\mathbf{F}$ of rate
\begin{align}
\nn \frac{1}{n}H(\mathbf{F}) 
&\leq \sum_{i=0}^{\lfloor (r-1)/2\rfloor} I(X\wedge U_{2i+1}\mid Y, U^{2i})
\\\label{el11}
&\quad +\sum_{i=1}^{\lfloor r/2\rfloor} I(Y\wedge U_{2i}\mid X, U^{2i-1})
+ \ep,
\end{align}
such that $(J, \mathbf{F})$ satisfies (\ref{el1}). For
$r=2$, such a construction was given by Ahlswede-Csisz{\'a}r
\cite[Theorem 4.4]{AhlCsi98}. (In their construction,
$\mathbf{F}$ was additionally a function of $J$.) The extension of
their construction to a general $r$ is straightforward, and is
relegated to the appendix.

It remains to prove (\ref{el5}).
Note
\begin{align}\nn
&I(X, Y \wedge U^r) - \sum_{i=0}^{\lfloor (r-1)/2\rfloor}
I(X\wedge U_{2i+1}\mid Y, U^{2i})
\\\nn &\hspace*{2cm} - \sum_{i=1}^{\lfloor
r/2\rfloor} I(Y\wedge U_{2i}\mid X, U^{2i-1})\\\label{el3}
&=\sum_{i=0}^{\lfloor (r-1)/2\rfloor} I(Y\wedge U_{2i+1}\mid
U^{2i}) + \sum_{i=1}^{\lfloor r/2\rfloor} I(X\wedge U_{2i}\mid
U^{2i-1}).
\end{align}
Further, from conditions (P1)-(P3) it follows that
\begin{align}\nn
&\sum_{i=0}^{\lfloor (r-1)/2\rfloor} I(Y\wedge U_{2i+1}\mid
U^{2i}) + \sum_{i=1}^{\lfloor r/2\rfloor} I(X\wedge U_{2i}\mid
U^{2i-1})\\\nn &\hspace*{5cm} - I(Y\wedge X)
\\\nn &=\sum_{i=1}^{\lfloor (r-1)/2\rfloor} I(Y\wedge U_{2i+1}\mid  U^{2i})
+ \sum_{i=2}^{\lfloor r/2\rfloor} I(X\wedge U_{2i}\mid U^{2i-1})
\\\nn&\qquad+I(X\wedge U_2\mid U_1) + I(Y\wedge U_1) -
I(Y\wedge X)
\\\nn &=\sum_{i=1}^{\lfloor (r-1)/2\rfloor} I(Y\wedge U_{2i+1}\mid  U^{2i})
+ \sum_{i=2}^{\lfloor r/2\rfloor} I(X\wedge U_{2i}\mid U^{2i-1})
\\\nn&\qquad+ I(X\wedge U_2\mid U_1) - I(X\wedge Y\mid U_1)
\\\nn
&=\sum_{i=1}^{\lfloor (r-1)/2\rfloor} I(Y\wedge U_{2i+1}\mid
U^{2i}) + \sum_{i=2}^{\lfloor r/2\rfloor} I(X\wedge U_{2i}\mid
U^{2i-1})
\\\nn &\hspace*{5cm} - I(X\wedge Y\mid U_1, U_2)\\\label{el4} &= ... =
-I(X\wedge Y\mid U^r) = 0.
\end{align}
Combining (\ref{el3}) and (\ref{el4}) we get (\ref{el5}).

\emph{Converse.} Let $R\geq 0$ be an achievable $r$-interactive CI
rate. Then, for all $0 < \ep <1$, for some $n\geq 1$, there exists
an $r$-interactive communication $\mathbf{F}$, and $\ep$-CR $J$
recoverable from $\mathbf{F}$, such that
 $(1/n)H(J, \mathbf{F})\leq R +\ep$ and $L = (J,\mathbf{F})$ satisfies (\ref{e_CIcond}).
Let $J$ be recovered as $J_1 = J_1(X^n, \mathbf{F})$ and $J_2 =
J_2(Y^n, \mathbf{F})$ by terminals
$\cX$ and
$\cY$, respectively, i.e., 
$\bPr{J = J_1 = J_2} \geq 1-\ep.$
 Further, let rv $T$
 be distributed uniformly over the set
 $\{1, ..., n\}$. Define rvs $U^r$ as follows:
\begin{align}
\nn U_1 &=  F_1, X^{T-1}, Y_{T+1}^n, T,\\
\nn U_i &= F_i, \qquad 2 \leq i < r, \\
\nn U_r &= \begin{cases}
(F_r, J_1), \quad r \text{ odd, }\\
(F_r, J_2), \quad r \text{ even. }
\end{cases}
\end{align}
We complete the proof for odd $r$; the proof for even $r$ can be
completed similarly. It was shown by Kaspi \cite[equations
(3.10)-(3.13)]{Kas85} that
\begin{align}
\nn &U_{2i+1} \mc X_T, U^{2i} \mc Y_T, \quad 0 \leq i \leq \lfloor(r-1)/2\rfloor, \\
\nn &U_{2i} \mc Y_T, U^{2i-1} \mc X_T, \quad 1 \leq i \leq \lfloor
r/2\rfloor.
\end{align}
Next, note from (\ref{ec3ff}) that
\begin{align}
\nn & \epsilon + \ep\log |\cX||\cY| + h(\ep) 
\\\nn&\geq \frac{1}{n}I(X^n \wedge Y^n \mid J_1, \mathbf{F}) \\
\nn &\geq \frac{1}{n}\sum_{i=1}^nI(X_i \wedge Y^n \mid X^{i-1},  J_1, \mathbf{F}) \\
\nn &\geq \frac{1}{n}\sum_{i=1}^nI(X_i \wedge Y_i \mid X^{i-1}, Y_{i+1}^n, J_1, \mathbf{F}) \\
\label{el6} &=I(X_T\wedge Y_T \mid U^r).
\end{align}
Similarly, it holds that
\begin{align}
\label{el7}\epsilon + \ep\log |\cX||\cY| + h(\ep) &\geq I(X_T
\wedge Y_{T+1}^n \mid X^{T-1}, J_1, \mathbf{F}, T) .
\end{align}
The entropy rate of $(J, \mathbf{F})$ is now bounded as
\begin{align}
\nn &\frac{1}{n}H(J, \mathbf{F}) 
\\\nn &\geq \frac{1}{n}H(J_1, \mathbf{F}) - \frac{1}{n}H(J_1\mid J)\\
\nn &\geq \frac{1}{n}H(J_1, \mathbf{F}) - \ep\log|\cX||\cY| - h(\ep)\\
\nn &= \frac{1}{n}I(X^n, Y^n \wedge J_1, \mathbf{F}) - \ep\log|\cX||\cY| - h(\ep)\\
\nn &= H(X_T, Y_T) - \frac{1}{n}H(X^n \mid J_1, \mathbf{F}) 
        \\\nn &\qquad  - \frac{1}{n}H(Y^n \mid X^n, J_1, \mathbf{F}) 
         - \ep\log|\cX||\cY| - h(\ep)\\
\nn &= H(X_T, Y_T) - H(X_T \mid X^{T-1}, J_1, \mathbf{F},T) 
                    \\\nn &\qquad  -
H(Y_T \mid X^{T-1},Y_{T+1}^n, X_T, X_{T+1}^n, J_1, \mathbf{F},T)  
\\\nn &\qquad - \ep\log|\cX||\cY| - h(\ep)                \\
\nn &\geq I(X_T, Y_T \wedge U^r) - \ep  -2 \ep\log|\cX||\cY|
-2 h(\ep),
\end{align}
where the second inequality follows from
Fano's inequality,
and the last inequality follows from (\ref{el7}). Consequently,
\begin{align}
R &\geq \frac{1}{n}H(J, \mathbf{F}) -\ep
\nn
\\&\label{el8} \geq I(X_T, Y_T \wedge
U^r) - 2(\ep  + \ep\log|\cX||\cY| + h(\ep)).
\end{align}
We now replace the rvs $U_1, ..., U_r$ with those taking values in
finites sets $\cU_1, ..., \cU_r$, respectively, with $\cU_1, ...,
\cU_r$ satisfying the cardinality bounds in condition (iii).
Similar bounds were derived in the context of interactive function
computation in \cite{MaIsh11}. For $1\leq l \leq r$, assume that 
rvs $\cU_1, ...,\cU_{l-1}$ satisfy the cardinality bounds. 
We consider odd $l$; the steps for even $l$ are similar. If the
rv $U_l$ does not satisfy the cardinality bound, from the Support
Lemma \cite[Lemma 15.4]{CsiKor11}, we can replace it with another
rv $\tilde{U}_l$ that takes less than or equal to
$|\cX|\prod_{i=1}^{l-1}|\cU_i| +1$ values, while keeping the
following quantities unchanged:
$$\bPP{X_TU^{l-1}}, \,\, I(X_T\wedge Y_T \mid U^r), \text{ and }  I(X_T, Y_T \wedge U^r).$$
Note that we have only altered $\bPP{U_l}$ in the joint
pmf $\bPP{X_TY_TU^r} = \bPP{U_l}\bPP{X_TU^{l-1}\mid
U_l}\bPP{Y_T\mid XU^{l-1}}$. Hence, the Markov relations in (P1)
remain unaltered. Furthermore, $\bPP{X_T Y_T} = \bPP{XY}$.
Finally, since the set of pmfs on a finite alphabet is compact, 
and the choice of $\ep$ above was
arbitrary, it follows upon taking $\ep \rightarrow 0$ in
(\ref{el6}) and (\ref{el8}) that there exists $U_1^r$ satisfying
(P1)-(P3) such that
$$R \geq I(X, Y \wedge U^r),$$
which completes the proof.\qed

\section{Can interaction reduce the communication rate?}\label{s_examples}
It is well known that the SK capacity can be attained by 
using a simple one-way communication from
terminal $\cX$ to terminal $\cY$ (or from $\cY$ to $\cX$).
Here we derive the minimum rate $R_{NI}$ of such 
noninteractive communication using the expression 
for $CI_i^r(X; Y)$ in (\ref{e_iCIr}). 
Since this expression has a 
\emph{double Markov structure}, it can be simplified by the
following observation (see \cite[Problem 16.25]{CsiKor11}): If rvs $U, X, Y$ satisfy
\begin{align}\label{e_doublemc}
U \mc X \mc Y, \quad X \mc U \mc Y,
\end{align}
then there exist functions $f= f(U)$ and $g=g(X)$ such that
\begin{enumerate}
\item[(i)] $\bPr{f(U) = g(X)} = 1$;

\item[(ii)] $X \mc g(X) \mc Y$.
\end{enumerate}
In particular, for rvs $U, X, Y$ that satisfy (\ref{e_doublemc}),
it follows from (i) above that
\begin{align}\nn
I(X, Y \wedge U) = I(X\wedge U) \geq I(g(X)\wedge f(U)) = H(g(X)).
\end{align}
Turning to (\ref{e_iCIr}), for rvs $U^r$ with $r$ odd, 
the observations above applied to the rvs $X$ and $Y$ 
conditioned on each realization $U^{r-1} = u^{r-1}$ implies that
there exists a function $g_1 = g_1\left(X, U^{r-1}\right)$ such that
\begin{align}\label{eh1}
X\mc g\left(X, U^{r-1}\right), U^{r-1} \mc Y,
\end{align}
and
$$I\left(X, Y \wedge U^r\right) \geq I\left(X, Y \wedge U^{r-1}\right) + H\left(g\left(X, U^{r-1}\right)\mid U^{r-1}\right),$$
where rv $U^{r-1}$ satisfies (P1), (P3). Similar observations hold for even $r$.
Thus, for the minimization in (\ref{e_iCIr}), conditioned on arbitrarily
chosen rvs $U^{r-1}$ satisfying (P1), (P3), the rv $U_r$ is
selected as a \emph{sufficient statistic for $Y$ given the observation
$X$} (sufficient statistic for $X$ given the observation $Y$) when
$r$ is odd ($r$ is even). Specifically, for $r=1$, we have
\begin{align}\label{eh2}
CI_i^1(X;Y) = \min_{X\mc g_1(X)\mc Y}H\left(g_1(X)\right),
\end{align}
and
\begin{align}\label{eh3}
CI_i^1(Y;X) = \min_{Y\mc g_2(Y)\mc X}H\left(g_2(Y)\right).
\end{align}
The answer to the optimization problems in (\ref{eh2}) and
(\ref{eh3}) can be given explicitly. In fact, we specify next a
minimal sufficient statistic for $Y$ on the basis of $X$.
Define an equivalence relation on $\cX$ as follows:
\begin{align}\label{eh3b}
x\sim x^\prime \Leftrightarrow \bP{Y\mid X}{y\mid x} =\bP{Y\mid
X}{y\mid x^\prime}, \quad y\in \cY.
\end{align}
Let $g_1^*$ be the function corresponding to the equivalence classes
of $\sim$. We claim that $g_1^*$ is a minimal sufficient statistic for $Y$
on the basis of $X$. This expression for the minimal sufficient statistic
was also given in \cite[Lemma 3.5(4)]{KamAna10}.
Specifically, $X \mc g_1^*(X) \mc Y$
since with $g_1^*(X) = c$, say, we have 
\begin{align}\nn
&\bP{Y\mid g_1^*(X)}{y\mid c}  
\\\nn &=\sum_{x\in \cX} \bP{Y, X \mid g_1^*(X) }{y, x \mid c}
\\\nn &= \sum_{x: g_1^*(x)=c} \bP{ X \mid g_1^*(X) }{ x \mid c}\bP{Y\mid X, g_1^*(X)}{y\mid x, c}
\\\nn &= \bP{Y\mid X, g_1^*(X)}{y\mid x, c}, \quad \forall \, x \text{ with } g_1^*(x) = c.
\end{align}
Also, if $g_1(X)$ satisfies $X \mc g_1(X) \mc Y$ then $g_1^*$ is a function
of $g_1$. To see this, let $g_1(x) = g_1(x^\prime) = c$ for some 
$x, x^\prime\in \cX$. Then, 
$$\bP{Y\mid g_1(X)}{y\mid c} = \bP{Y\mid X}{y\mid x} = \bP{Y\mid X}{y\mid x^\prime}, \quad y \in \cY,$$
so that $g_1^*(x) = g_1^*(x^\prime)$. Since $g_1^*$ is a minimal sufficient statistic for $Y$ on the basis of $X$, it follows from 
(\ref{eh2}) that
$$CI_i^1(X; Y) = H\left(g_1^*(X)\right),$$
and similarly, with $g_2^*(Y)$ defined analogously,
$$CI_i^1(Y; X) = H\left(g_2^*(Y)\right).$$
Therefore, from (\ref{e_main2}), the minimum rate $R_{NI}$ 
of a noninteractive communication for generating a maximum
rate SK is given by 
\begin{align}\label{e:nonint_comm}
R_{NI} = \min\left\{H\left(g_1^*(X)\right), H\left(g_1^*(X)\right)\right\} - I(X\wedge Y).
\end{align}

From the expression for $R_{NI}$, it is clear that the rate of 
noninteractive communication
can be reduced by replacing $X$ and $Y$ with their respective minimal 
sufficient statistics $g_1^*(X)$ and $g_2^*(Y)$.
Can the rate of
communication required for generating an optimum rate SK be
reduced by resorting to complex interactive communication protocols
defined in Section \ref{s_def}? 
To answer this question we must compare the expression for 
$R_{NI}$ with $R_{SK}$. Specifically, from Theorem \ref{t_main}
and the Corollary following it, interaction reduces the rate of communication
iff, for some $r>1$,
\begin{align}\label{e:interaction_helps?}
CI_i^r(X; Y) < \min\left\{H\left(g_1^*(X)\right), H\left(g_1^*(X)\right)\right\},
\end{align}
where $g_1^*$ and $g_2^*$ are as in (\ref{e:nonint_comm}); interaction does not
help iff
$$CI_i(X\wedge Y) = \min\left\{H\left(g_1^*(X)\right), H\left(g_1^*(X)\right)\right\}.$$
Note that instead of comparing with 
$CI_i^r(X; Y)$ in (\ref{e:interaction_helps?}), we can also compare with
$CI_i^r(Y;X)$.

We shall
explore this question here, and give an example where the answer
is in the affirmative. In fact,
we first show that interaction does not help in the case of binary
symmetric sources. Then we give an example where interaction does
help.

\subsection{Binary Symmetric Sources}\label{s_BS}
For binary rvs $X$ and $Y$, we note a property of rvs $U^r$ that
satisfy the conditions (P1)-(P3) in Theorem \ref{t_char}.

\begin{lemma}\label{l_bin}
Let $X$ and $Y$ be $\{0,1\}$ valued rvs with $I(X\wedge Y)\neq 0$.
Then, for rvs $U_1,...,U_r$ that satisfy the conditions (P1)-(P3)
in Theorem \ref{t_char}, for every realization $u_1,..., u_r$ of
$U_1, ..., U_r$, one of the following holds:
\begin{align}\label{ee5}
H(X\mid U^r=u^r) = 0,\,\, \text{ or }\,\, H(Y\mid U^r=u^r) = 0.
\end{align}
\end{lemma}
{\it Proof.} Given a sequence $u^r$, assume that
$$H(X\mid U^r=u^r) > 0 \,\,\text { and } \,\,H(Y\mid U^r=u^r) > 0,$$
which is equivalent to
\begin{align}
\nn \bP{X\mid U^r}{1\mid u^r}\bP{X\mid U^r}{0\mid u^r} &>0
 \,\,\text{ and }
\\\label{ee1} \bP{Y\mid U^r}{1\mid u^r}\bP{Y\mid U^r}{0\mid u^r} &>0.
\end{align}
We consider the case when $r$ is even; the case of odd $r$ is
handled similarly. From the Markov conditions $X\mc U^r\mc Y$ and
$X\mc Y, U^{r-1}\mc U_r$, we have
\begin{align}\nn
&\bP{X,Y\mid U^r}{x, y\mid u^r}
\\\nn & = \bP{X\mid U^r}{x\mid
u^r}\bP{Y\mid U^r}{y\mid u^r}\\\nn & = \bP{X\mid Y,U^{r-1}}{x\mid
y, u^{r-1}}\bP{Y\mid U^r}{y\mid u^r}, \quad x,y \in \{0,1\}.
\end{align}
Since $\bP{Y\mid U^r}{y\mid u^r}> 0$ from (\ref{ee1}), we have
\begin{align}\nn
\bP{X\mid U^r}{x\mid u^r} = \bP{X\mid Y,U^{r-1}}{x\mid y,
u^{r-1}}, \quad x,y \in \{0,1\},
\end{align}
which further implies
\begin{align}\nn
\bP{X\mid Y,U^{r-1}}{x\mid 1, u^{r-1}} &= \bP{X\mid
Y,U^{r-1}}{x\mid 0, u^{r-1}}, \\\nn &\hspace*{3cm} x \in \{0,1\}.
\end{align}
Hence, $I(X\wedge Y\mid U^{r-1} = u^{r-1} )= 0$. Noting from
(\ref{ee1}) that
$$\bP{X\mid U^{r-1}}{1\mid u^{r-1}}\bP{X\mid U^{r-1}}{0\mid u^{r-1}}>0,$$
we can do the same analysis as above, again for $r-1$. Upon
repeating this process $r$ times we get $I(X\wedge Y)=0$, which is
a contradiction. Therefore, either $H(X\mid U^r=u^r) = 0$ or
$H(Y\mid U^r=u^r) = 0$ holds.\qed

Note that
$$CI_i^r(X;Y) = H(X,Y) - \max_{U^r} H(X, Y \mid U^r),$$
where the $\max$ is taken over rvs $U^r$ as in Theorem
\ref{t_char}. If $H(X\mid U^i = u^i) = 0$, it follows that
\begin{align}\nn
I\left(X\wedge Y\mid U^i = u^i\right)&=0, \text{ and
}\\\nn H(X,Y \mid U^i=u^i, U_{i+1}^r) &= H(Y \mid U^i=u^i,
U_{i+1}^r) 
\\\label{ef1}&\leq H(Y \mid U^i=u^i).
\end{align}
Similarly, $H(Y\mid U^i = u^i) = 0$ implies
\begin{align}\nn
I\left(X\wedge Y\mid U^i = u^i\right)&=0, \text{ and
}\\\label{ef2} H(X,Y \mid U^i=u^i, U_{i+1}^r)  &\leq H(X \mid
U^i=u^i).
\end{align}
For a sequence $u^r$ with $\bP{U^r}{u^r}> 0$, let $\tau(u^r)$ be
the minimum value of $i$ such that
$$H(X\mid U^i = u^i) = 0 \text{ or } H(Y\mid U^i = u^i)=0;$$
if $X$ and $Y$ are independent, $\tau(u^r)=0$. Note that $\tau$ is
a stopping-time adapted to $U_1 ,..., U_r$. Then, from
(\ref{ef1}), (\ref{ef2}), $CI_i^r(X;Y)$ remains unchanged if we
restrict the support of $U^r$ to sequences $u^r$ with $u_i =\phi$
for all $i> \tau(u^r)$. Furthermore, the Markov condition (P1)
implies that if for a sequence $u^r$, $\tau = \tau(u^r)$ is odd
then
$$\bP{Y \mid X, U^\tau}{y \mid x, u^\tau} = \bP{Y \mid X, U^{\tau-1}}{y \mid x, u^{\tau-1}},$$
and so if
$$\bP{X\mid U^\tau}{1\mid u^\tau}\bP{X\mid U^\tau}{0\mid u^\tau}>0,$$
it holds from the definition of $\tau$ that
$$\bP{Y\mid U^\tau}{1\mid u^\tau}\bP{Y\mid U^\tau}{0\mid u^\tau}>0,$$
which is a contradiction. Therefore, we have $H(X\mid U^\tau =
u^\tau) = 0$. Similarly, $H(Y\mid U^\tau = u^\tau) = 0$ holds for
even $\tau$. To summarize,
\begin{align}\label{ef3}
CI_i^r(X;Y) = \min_{U^\tau} I\left(X,Y \wedge U^\tau\right),
\end{align}
where $U^r$ are rvs satisfying (P1)-(P3), and $\tau$ is the
stopping-time defined above.

We show next that for binary
symmetric sources, interaction can never reduce the rate of
communication for optimum rate SK generation. In fact, {\it 
we conjecture that for any binary rvs $X, Y$, $R_{NI} = R_{SK}$.}
\begin{theorem}\label{t_BS}
Let $X$ and $Y$ be $\{0,1\}$-valued rvs, with
\begin{align}\nn
&\bPr{X=0, Y=0} = \bPr{X=1, Y=1} = \frac{1}{2}(1-\delta),\\
\label{e_BS} &\bPr{X=0, Y=1} = \bPr{X=1, Y=0} =
\frac{1}{2}\delta,\quad 0<\delta<\frac{1}{2}.
\end{align}
Then,
$$CI_i(X\wedge Y) = \min\{H(X); H(Y)\},$$
i.e., interaction does not help to reduce the communication
required for optimum rate SK generation.
\end{theorem}
\begin{remark*}
As a consequence of Theorem \ref{t_BS}, for sources with joint
distribution as in (\ref{e_BS}), the second inequality in
(\ref{e_CIineq}) can be strict. Specifically, it was noted by
Wyner (see the discussion following equation (1.19) in
\cite{Wyn75}) that for binary symmetric sources, $CI_W(X\wedge Y) <
1$. From Theorem \ref{t_BS}, we have
$$CI_i(X\wedge Y) =\min\{H(X); H(Y)\} = 1.$$
Thus, for such sources, $CI_W(X\wedge Y) < CI_i(X\wedge Y)$.
\end{remark*}

\noindent{\it Proof.} Denote by $\cU^r_0$ the following set of
stopped sequences in $\cU^r$:\\
For $i\leq r$, for a sequence $u^r\in \cU^r$ the stopped sequence
$u^i \in \cU^r_0$ if:
\begin{align}\nn
&H\left(X\mid U^j = u^j\right)>0, H\left(Y\mid U^j =
u^j\right)>0,\quad \forall\, j<i, \text{ and }\\\nn
&H\left(X\mid U^i = u^i\right)=0 \text{ or } H\left(Y\mid U^i = u^i\right)=0.
\end{align}
For $i \in\{0,1\}$, define the following subsets of $\cU_0^r$:
\begin{align}\nn
\nn \cU_i^X &= \left\{u^\tau \in \cU_0^r : \tau\text{ is odd},
\bP{X\mid U^\tau}{i\mid u^\tau} = 1\right\}, 
\\\nn \cU_i^Y &=
\left\{u^\tau \in \cU_0^r : \tau\text{ is even},\bP{Y\mid
U^\tau}{i\mid u^\tau} = 1\right\}.
\end{align}
By their definition the sets $\cU_0^X, \cU_1^X, \cU_0^Y,$ and
$\cU_1^Y$ are disjoint, whereby we have
\begin{align}\nn
\bP{U^\tau}{\cU_0^r} &= \bP{U^r}{\cU_0^X \bigcup \cU_1^X \bigcup
\cU_0^Y \bigcup \cU_1^Y} \\\label{ee11} &= \sum_{i=0}^1
\left[\bP{U^r}{\cU^X_i} + \bP{U^r}{\cU^Y_i}\right]= 1.
\end{align}
For $u^\tau \in \cU_0^r$, denote by $p(u^\tau)$ the probability
$\bP{U^\tau}{u^\tau}$. Further, for $u^\tau \in \cU_0^X\bigcup
\cU_1^X$, denote by $W^{u^\tau}:\cX \rightarrow \cY$ the
stochastic matrix corresponding to $\bP{Y\mid X,
U^\tau}{\cdot\mid\cdot, u^\tau}$, and for $u^\tau \in
\cU_0^Y\bigcup \cU_1^Y$, denote by $T^{u^\tau}:\cY \rightarrow
\cX$ the stochastic matrix corresponding to $\bP{X\mid Y,
U^\tau}{\cdot\mid\cdot, u^\tau}$. With this notation, the
following holds:
\begin{align}\nn
&\frac{1}{2}(1-\delta) 
\\\nn &= \bP{X,Y}{i,i} 
\\\nn  &= \sum_{u^\tau \in
\cU_i^X}p(u^\tau)W^{u^\tau}(i \mid i) + \sum_{u^\tau \in
\cU_i^Y}p(u^\tau)T^{u^\tau}(i \mid i),
\\\label{ee2} &\hspace*{6.3cm} i \in \{0,1\},
\end{align}
since the sets $\cU_0^X, \cU_1^X, \cU_0^Y,\cU_1^Y$ are disjoint.
Upon adding (\ref{ee2}) for $i =0, 1$, we get
\begin{align}\nn
&\sum_{i=0}^1\left[\sum_{u^\tau \in
\cU_i^X}p(u^\tau)W^{u^\tau}(i \mid i) + \sum_{u^\tau \in
\cU_i^Y}p(u^\tau)T^{u^\tau}(i \mid i)\right]
\\\nn &=(1-\delta).
\end{align}
Furthermore, from (\ref{ee11}) we get
\begin{align}\nn
1 &= \sum_{i=0}^1\sum_{u^\tau \in \cU_i^X}p(u^\tau) + \sum_{u^\tau
\in \cU_i^Y}p(u^\tau).
\end{align}
Therefore, since the function $g(z) = -z\log z$ is concave for $0<
z < 1$, the Jensen's inequality yields
\begin{align}\nn
g(1-\delta) &\geq \sum_{i=0}^1\sum_{u^\tau \in
\cU_i^X}p(u^\tau)g\left(W^{u^\tau}(i \mid i)\right) + 
\\\label{ee3}
&\hspace*{2cm}
\sum_{u^\tau
\in \cU_i^Y}p(u^\tau)g\left(T^{u^\tau}(i \mid i)\right)
\end{align}
Similarly, using
\begin{align}\nn
\frac{1}{2}\delta &=\bP{X,Y}{i,j}
\\\nn &= \sum_{u^\tau \in
\cU_i^X}p(u^\tau)\left(1- W^{u^\tau}(i \mid i)\right) +
\\\nn &\quad \sum_{u^\tau \in \cU_j^Y}p(u^\tau)\left( 1- T^{u^\tau}(j \mid
j)\right),\quad i\neq j, i,j \in \{0,1\},
\end{align}
we get
\begin{align}\nn
 g(\delta) &\geq \sum_{i=0}^1\sum_{u^\tau \in
\cU_i^X}p(u^\tau)g\left(1 - W^{u^\tau}(i \mid i)\right) +
\\ &\hspace*{2cm}\sum_{u^\tau \in \cU_i^Y}p(u^\tau)g\left(1- T^{u^\tau}(i \mid
i)\right)
\label{ee4}
\end{align}
On adding (\ref{ee3}) and (\ref{ee4}) we get
\begin{align}\nn
h(\delta) &= g(\delta) + g(1-\delta) 
\\\nn &\geq \sum_{i=0}^1\sum_{u^\tau
\in \cU_i^X}p(u^\tau)h\left(W^{u^\tau}(i \mid i)\right) +
\\\nn &\qquad \sum_{u^\tau \in \cU_i^Y}p(u^\tau)h\left(T^{u^\tau}(i \mid
i)\right),
\end{align}
where $h$ is the binary entropy function. Note that the right side
above equals $H(X, Y\mid U^\tau)$, which yields
$$h(\delta) = \max\{H(X\mid Y); H(Y\mid X)\} \geq H(X, Y\mid U^\tau).$$
Since rvs $U^r$ above were arbitrary, we have from (\ref{ef3}),
\begin{align}
\nn
CI_i^r(X;Y) &\geq H(X, Y) - \max\{H(X\mid Y); H(Y\mid X\}) 
\\\nn &= \min\{H(X); H(Y)\}  .
\end{align}

Combining this with (\ref{e_CIineq}), we obtain
$$CI_i^r(X;Y)= \min\{H(X); H(Y)\}.$$\qed

\subsection{An example where interaction does help}
Consider rvs $X$ and $Y$ with $\cX =\cY =\{0,1,2\}$, and with
joint pmf:
\begin{align}\nn
\left[\begin{matrix} a &a &a\\b &a &a\\a &c &a
\end{matrix}\right],
\end{align}
where $a, b, c$ are nonnegative, $7a+b+c =1$, and $c\neq a$, which
holds iff $b\neq 1- 8a$. Assume that
\begin{align}\label{eg2}
2a>b > a.
\end{align}
From (\ref{e:interaction_helps?}), to show that interaction helps, it suffices to find rvs $U_1, ..., U_r$
satisfying (P1)-(P3) such that
\begin{align}\label{e:ex2_nonint_comm}
I\left(X,Y \wedge U_1, ..., U_r\right) < \min\left\{H\left(g_1^*(X)\right), H\left(g_2^*(Y)\right)\right\},
\end{align}
where $g_1^*$ and $g_2^*$ are as in (\ref{e:nonint_comm}). From (\ref{eh3b}),
$g_1^*(x) = g_1^*(x^\prime)$ iff
\begin{align}
\frac{\bP{Y,X}{y,x}}{\bP{Y,X}{y,x^\prime}} = \frac{\bP{X}{x}}{\bP{X}{x^\prime}},\quad y \in \cY,
\end{align}
i.e., the ratio $\frac{\bP{Y,X}{y,x}}{\bP{Y,X}{y,x^\prime}}$ does not depend on 
$y$. Therefore, for the pmf above, $g_1^*(X)$ and $g_2^*(Y)$ are equivalent
to $X$ and $Y$, respectively. Thus,
$$ \min\left\{H\left(g_1^*(X)\right), H\left(g_2^*(Y)\right)\right\} = \min\{H(X); H(Y)\},$$
where $H(X) = H(Y)$ for the given pmf.

Next, let $U_1 = f_1(X)$, $U_2=f_2(Y, f_1(X))$, where $f_1$ and $f_2$
are given below:
\begin{align}\nn
f_1(x) &= \begin{cases} 1,\quad x=2,\\
2,\quad x=0,1,
\end{cases}
\\\nn
f_2(y,1)= 0,\forall\, y\in \{0,1,2\}, &\text{ and }\,\, f_2(y,2)= \begin{cases} 1,\quad y=0,\\
2,\quad y=1,2.
\end{cases}
\end{align}
Clearly, $U_1$ and $U_2$ satisfy (P1) and (P2).
For (P3), note that 
if $(U_1,U_2) = (1,0)$, then $X=2$, and
if $(U_1,U_2) = (2,1)$, then $Y=0$. Finally, if $(U_1,U_2) = (2,2)$, then
$X \in \{0,1\}$ and $Y \in \{1,2\}$, implying
\begin{align}
\nn \bP{X,Y\mid U_1, U_2}{x, y\mid 2,2} &=
\frac{\bP{X, Y}{x, y}}{4a}
\\\nn &=\frac{1}{4},
 \quad \forall\,\, (x,y)
\in\{0,1\}\times\{1,2\}.
\end{align}
Therefore, $I(X\wedge Y\mid U_1, U_2) =0$,
and so $U_1, U_2$ satisfy (P3). We show that (\ref{e:ex2_nonint_comm})
holds for this choice of $U_1, U_2$.
Specifically, $I\left(X,Y \wedge U_1, U_2\right) = H\left(U_1, U_2\right)$, and the
following holds:
\begin{align}\nn
&H(Y) - H\left(U_1, U_2\right)
\\\nn &= H(X) - H\left(U_1, U_2\right)
\\\nn &= H\left(X | U_1\right) - H\left(U_2| U_1\right)
\\\nn &= \bPr{f_1(X) = 2}\bigg[ H\left(X | f_1(X) = 2\right)
\\\nn &\hspace*{4cm} - H\left(f_2(2, Y)| f_1(X) = 2\right)\bigg]
\\\nn 
&=(5a+b)\left[h\left(\bP{X | f_1(X)}{0|2}\right) -  h\left(\bP{Y | f_1(X)}{0|2}\right)\right]
\\\nn &= (5a+b)
\left[h\left(\frac{3a}{5a+b}\right) -
h\left(\frac{a+b}{5a+b}\right)\right].
\end{align}
Then, from (\ref{eg2}),
$$\frac{a+b}{5a+b} < \frac{3a}{5a+b} < \frac{1}{2},$$
which implies (\ref{e:ex2_nonint_comm}) for $U_1, U_2$.

\section{Sufficient statistics and common information quantities}\label{s_suff}

In this work we encountered three CI quantities: 
Shannon's mutual information $I(X\wedge Y)$, Wyner's CI 
$CI_W(X\wedge Y)$, and interactive CI 
$CI_i(X\wedge Y)$. In fact, the first notion of CI
was given by G{\'a}cs and K{\"o}rner in the seminal work \cite{GacKor73}.
In particular, they specified the maximal common function of $X$ and $Y$, 
denoted here as $\mathtt{mcf}(X, Y)$, such that any other common function 
of $X$ and $Y$ is a function of $\mathtt{mcf}(X, Y)$; the G{\'a}cs-K\"orner
CI is given by $H(\mathtt{mcf}(X, Y))$. The following 
inequality ensues (see \cite{GacKor73,Wyn75}, and inequality (\ref{e_CIineq})):
$$H(\mathtt{mcf}(X, Y)) \leq I(X\wedge Y)\leq CI_W(X\wedge Y) \leq CI_i(X\wedge Y).$$

Since any good notion of CI between rvs $X$ and $Y$
measures the correlation between $X$ and $Y$, it is reasonable to expect
the CI to remain unchanged if
$X$ and $Y$ are replaced by their respective sufficient statistics.
The following theorem establishes this for the quantities
$H(\mathtt{mcf}(X, Y)), I(X\wedge Y), CI_W(X\wedge Y),$ and $H(\mathtt{mcf}(X, Y))$.

\begin{theorem}\label{t_suff}
For rvs $X$ and $Y$, let functions $g_1$ of $X$ and $g_2$ of $Y$
be such that $X\mc g_1(X) \mc Y$ and $X\mc g_2(Y)\mc Y$. Then the
following relations hold:
\begin{align}\nn
H(\mathtt{mcf}(X, Y)) &=
H\left(\mathtt{mcf}\left(g_1(X),g_2(Y)\right)\right),\\\nn
I(X\wedge Y) &= I\left(g_1(X)\wedge g_2(Y)\right),\\\nn CI_W(X\wedge
Y) &=CI\left(g_1(X)\wedge g_2(Y)\right),\\\nn CI_i^r(X; Y) &=
CI_i^r\left(g_1(X); g_2(Y)\right), \quad r\geq 1,\\\nn
CI_i(X\wedge Y) &= CI_i\left(g_1(X)\wedge g_2(Y)\right).
\end{align}
\end{theorem}
\begin{remark*}
(i) Theorem \ref{t_suff} implies that the minimum rate of communication 
for generating a maximum rate secret key remains unchanged if $X$
and $Y$ are replaced by $g_1(X)$ and $g_2(Y)$ as above, respectively. 

\noindent(ii) Note that $g_1(X)$ and $g_2(Y)$ above are, respectively, functions of 
$g_1^*(X)$ and $g_2^*(Y)$ defined through (\ref{eh3b}).
\end{remark*}
{\bf Proof.} First note that 
\begin{align}\label{eh5}
I(X\wedge Y) = I\left(g_1(X)\wedge Y\right) =I\left(g_1(X)\wedge
g_2(Y)\right).
\end{align}

Next, consider the interactive CI. 
From (\ref{eh5}), any protocol that
generates an optimum rate SK for the sources $g_1(X)$ and
$g_2(Y)$ also generates an optimum rate SK for the sources $X$
and $Y$. Thus, the minimum communication rate for prior
protocols is bounded below by the minimum communication rate for
the latter protocols, so that by Theorem \ref{t_main},
\begin{align}
\nn
&CI_i^r\left(g_1(X); g_2(Y)\right) - I\left(g_1(X)\wedge g_2(Y)\right)
\\\nn &\geq CI_i^r\left(X; Y\right) - I\left(X\wedge Y\right),
\end{align}
which,
by (\ref{eh5}), is
\begin{align}\label{eh6}
CI_i^r\left(g_1(X); g_2(Y)\right) \geq CI_i^r\left(X; Y\right) .
\end{align}
In fact, (\ref{eh6}) holds with equality: We claim that
any choice of rvs $U^r$ that satisfy (P1)-(P3) also
satisfy the following Markov relations:
\begin{align}\nn
& U_{2i+1} \mc g_1(X), U^{2i} \mc g_2(Y), \quad 0 \leq i \leq
\lfloor(r-1)/2 \rfloor,
\\\nn& U_{2i} \mc g_2(Y), U^{2i-1}
\mc g_1(X),\quad 1 \leq i \leq \lfloor r/2 \rfloor,
\\\label{eh7} &g_1(X)
\mc  U^{r} \mc g_2(Y).
\end{align}
It follows that
\begin{align}\nn
CI_i^r\left(g_1(X); g_2(Y)\right) &\leq I\left(g_1(X), g_2(Y) \wedge U^r\right)
\\\nn &\leq I\left(X, Y \wedge U^r\right),
\end{align}
and consequently,
$$CI_i^r\left(g_1(X); g_2(Y)\right) \leq CI_i^r\left(X; Y\right).$$
Thus, by (\ref{eh6}),
\begin{align}\label{eh7b}
CI_i^r\left(g_1(X); g_2(Y)\right) = CI_i^r\left(X; Y\right).
\end{align}
Taking the limit $r\rightarrow \infty$ we get
$$CI_i\left(g_1(X)\wedge g_2(Y)\right) = CI_i\left(X\wedge Y\right).$$
It remains to establish (\ref{eh7}); instead, using induction 
we establish the following stronger Markov relations: For $1\leq i \leq r$,
\begin{align}\nn
& U_{i} \mc g_1(X), U^{i-1} \mc Y, \quad i \text{ odd},
\\\nn & U_{i} \mc g_2(Y), U^{i-1}
\mc X,\quad i \text{ even},
\\\label{eh8}
&X \mc  g_1(X),U^i \mc Y \text{ and } X \mc  g_2(Y),U^i \mc Y.
\end{align}
Clearly, (\ref{eh8}) implies the first two Markov relations in (\ref{eh7}).
The last Markov chain in (\ref{eh7}) follows upon observing
\begin{align}\nn
0 = I\left(X\wedge Y\mid U^r\right) \geq I\left(g_1(X)\wedge
g_2(Y)\mid U^r\right).
\end{align}
To see that (\ref{eh8}) holds for $i=1$ note that
\begin{align}\nonumber
&I\left(X\wedge Y\mid g_1(X), U_1\right) 
\\\nn &\leq I\left(X\wedge Y\mid g_1(X)\right)  + I\left(U_1 \wedge Y\mid g_1(X), X\right) = 0,
\end{align}
and
\begin{align}
\nonumber 
&I\left(X\wedge Y\mid g_2(Y), U_1\right) 
\\\nn &\leq I\left(X\wedge Y\mid g_2(Y)\right) + I\left(U_1 \wedge Y, g_2(Y)\mid X\right) = 0.
\end{align}
Next, assume that (\ref{eh8}) holds for an even $i$.
%
Then, from (P1) we get:
\begin{align}\nn
&I\left(Y \wedge U_{i+1}\mid X, U^{i}\right) = 0 \\\nn
\Leftrightarrow &I\left(Y \wedge U_{i+1}\mid X, g_1(X),
U^{i}\right) = 0\\\nn \Leftrightarrow &I\left(Y \wedge X,
U_{i+1}\mid g_1(X), U^{i}\right) = I\left(Y \wedge X\mid g_1(X),
U^{i}\right)=0,
\end{align}
where the last equality follows from (\ref{eh8}). 
From the last inequality above we
have
$$U_{i+1} \mc g_1(X), U^i \mc Y \,\,\text{ and }\,\,X \mc g_1(X), U^{i+1} \mc Y.$$
Furthermore, it also follows from (\ref{eh8}) that
\begin{align}\nn
I\left(X \wedge Y\mid g_2(Y), U^{i+1}\right) &\leq I\left(X,
U_{i+1} \wedge Y\mid g_2(Y), U^{i}\right)
 \\\nn
&= I\left(U_{i+1} \wedge Y\mid g_2(Y), X, U^{i}\right)\\\nn &\leq
I\left(U_{i+1} \wedge Y\mid  X, U^{i}\right) =0,
\end{align}
where the last equality follows from (P1). Thus, we have
$$X \mc g_2(Y), U^{i+1} \mc Y,$$
establishing the validity of (\ref{eh8}) for $i+1$.
The proof of (\ref{eh7}) can be completed by induction
by using a similar argument for odd $i$.

Next, we consider the G{\'a}cs-K{\"o}rner CI. Note that any common function of 
$g_1(X)$ and $g_2(Y)$ is also a
common function of $X$ and $Y$. Consequently,
\begin{align}\label{eh9}
H(\mathtt{mcf}(X, Y)) \geq H(\mathtt{mcf}(g_1(X), g_2(Y))).
\end{align}
For the reverse inequality, observe that for an rv $U$ 
such that $H(U|Y) = H(U|X) = 0$
we have
$$U \mc X \mc g_1(X) \mc Y.$$
Thus, $H\left(U| g_1(X)\right) \leq H(U | Y) = 0,$
and similarly, $H\left(U| g_2(Y)\right) = 0.$
In particular, it holds that
$$H\left(\mathtt{mcf}(X, Y)| g_1(X)\right) = H\left(\mathtt{mcf}(X, Y)| g_2(Y)\right) = 0,$$
and so,
$$H(\mathtt{mcf}(X, Y)) \leq H(\mathtt{mcf}(g_1(X), g_2(Y))),$$
which along with (\ref{eh9}) yields
$$H(\mathtt{mcf}(X, Y)) = H(\mathtt{mcf}(g_1(X), g_2(Y))).$$

Finally, we consider Wyner's CI and claim that this, too, remains
unchanged upon replacing the sources with their respective sufficient
statistics (for the other source). It suffices to show that
$$CI_W(X\wedge Y) = CI_W(g(X)\wedge Y),$$
for a function $g$ such that $X \mc g(X)\mc Y$. Consider an rv $W$
for which $X\mc W\mc Y$ is satisfied. We have
$$0 = I(X\wedge Y\mid W)\geq I\left(g(X)\wedge Y\mid W\right).$$
It follows from (\ref{e_CIchar}) that
\begin{align}\label{ei1}
CI_W(X\wedge Y) \geq CI_W\left(g(X)\wedge Y\right).
\end{align}
On the other hand, for an rv $L = L\left(g^n\left(X^n\right), Y^n\right)$
we have 
$$\frac{1}{n}I\left(X^n\wedge Y^n\mid L\right) = \frac{1}{n}I\left(g^n\left(X^n\right)\wedge Y^n\mid L\right),$$
since
\begin{align}\nn
I\left(X^n\wedge Y^n\mid L, g^n\left(X^n\right)\right) 
&\leq I\left(X^n\wedge Y^n, L\mid  g^n\left(X^n\right)\right) 
\\&= I\left(X^n\wedge Y^n\mid g^n\left(X^n\right)\right) = 0.
\nonumber
\end{align}

Thus, from the definition of $CI_W(g(X)\wedge Y)$
we get 
$$CI_W(X\wedge Y) \leq CI_W(g(X)\wedge Y),$$
so that, by (\ref{ei1}),
$$CI_W(X\wedge Y) = CI_W(g(X)\wedge Y).$$\qed


\section{Discussion}
\subsection{Local Randomization}\label{s_discA}
Although independent local randomization was not
allowed in our formulation, our main result characterizing 
$R_{SK}$ holds even when 
such randomization is available. Consider a model where
terminals $\cX$ and $\cY$, in additional to their respective
observations $X^n$ and $Y^n$, have access to finite-valued\footnote{The cardinalities of the
range spaces of $T_1$ and $T_2$ are allowed to be at
most exponential in $n$.}
rvs $T_1$ and $T_2$, respectively. The rvs $T_1$, $T_2$,
and $(X^n, Y^n)$ are mutually independent. The SK capacity is
defined as before, with $X^n$ and $Y^n$ now replaced by $(X^n, T_1)$
and $(Y^n, T_2)$, respectively. It is known \cite{Mau93, AhlCsi98}
that even with randomization the SK capacity equals $I(X\wedge Y)$.
For this model, denote the minimum rate of $r$-interactive communication required
to generate an SK of rate $I(X\wedge Y)$ by $\tilde{R}_{SK}^r$. 
\begin{lemma}\label{l:randomize}
For $r\geq 1$,
\begin{align}
\tilde{R}_{SK}^r = {R_{SK}^r}.
\nn
\end{align}
\end{lemma}
To see this, we define quantities $\tilde{R}_{CI}^r$ and $\tilde{CI}^r_i$
analogously to ${R}_{SK}^r$ and ${CI_{i}^r}$, with $X^n$ and $Y^n$ replaced by $(X^n, T_1)$
and $(Y^n, T_2)$, respectively. Note that this substitution is made even in condition (\ref{e_CIcond}),
i.e., the CR $J$ and the communication $\bF$ now are required to satisfy:
\begin{align}\label{e_CIcond2}
\frac{1}{n}I\left(X^n, T_1 \wedge Y^n, T_2 \mid J, \bF\right) \leq \ep.
\end{align}
We observe that (\ref{e_eq}) still holds, with $(X^n, T_1)$
and $(Y^n, T_2)$ replacing, respectively, $X^n$ and $Y^n$
on the right-side. Therefore, the proof of Theorem \ref{t_main}
is valid, and we get:
\begin{align}\label{eg1}
\tilde{R}_{CI}^r = \tilde{R}_{SK}^r = \tilde{CI}_{i}^r - I(X\wedge Y).
\end{align}
By its definition $\tilde{R}_{CI}^r \leq {R_{CI}^r}$, since $L = (J, \bF) = L(X^n, Y^n)$ satisfying
(\ref{e_CIcond}) will meet (\ref{e_CIcond2}) as well. We claim that 
$\tilde{R}_{CI}^r \geq {R_{CI}^r}$, which by (\ref{eg1}) and Theorem \ref{t_main}
implies Lemma \ref{l:randomize}. Indeed, consider CR $J$ recoverable from $\bF$
such that $(J, \bF)$ attain $\tilde{R}_{CI}^r$. Then, the condition (\ref{e_CIcond2})
gives
\begin{align}\nn
\frac{1}{n}I\left(X^n \wedge Y^n \mid J, \bF, T_1, T_2\right) \approx 0.
\end{align}
So, there exist $t_1, t_2$ such that
conditioned on $T_1 = t_1, T_2 = t_2$  the CR $J$ is still recoverable from 
$\bF$, and
\begin{align}\nn
\frac{1}{n}I\left(X^n \wedge Y^n \mid J, \bF, T_1 = t_1, T_2 = t_2\right) \approx 0.
\end{align}
Thus, with $T_1 = t_1, T_2 = t_2$ fixed, $(J, \bF)$ constitutes
a feasible choice in the definition of 
$R_{CI}^r$. 
Since the number of values taken by $\bF$ can only decrease
upon fixing $T_1 = t_1, T_2 =t_2$, we get $\tilde{R}_{CI}^r \geq {R}_{CI}^r$.
Therefore, the availability of local randomization does not decrease the rate 
of communication required for generating an optimum rate SK.

\subsection{Less-than-optimum rate SKs}
SK generation is linked intrinsically to the efficient generation of CR. 
For $\rho \geq 0$, a rate $R\geq 0$ is an achievable CR rate for
$\rho$ if for every $0< \ep<1$ there exists, for some $n \geq 1$, an 
$\ep$-CR $L$ with
$$\frac{1}{n}H(L) \geq R -\ep,$$
recoverable from an $r$-interactive communication $\mathbf{F}$,  for 
arbitrary $r$, of 
rate $$\frac{1}{n}H(\mathbf{F}) \leq \rho+ \ep;$$ the maximum 
achievable CR rate for $\rho$ is denoted by $CR(\rho)$.
Similarly, denote by $C(\rho)$ the maximum rate of an SK
that can be generated using a communication as above.
It can be shown in a straightforward manner that 
\begin{align}\label{e:SKandCR}
C(\rho) = CR(\rho) - \rho.
\end{align}

\begin{figure}
\vspace{.5cm}
\begin{center}
\epsfig{file=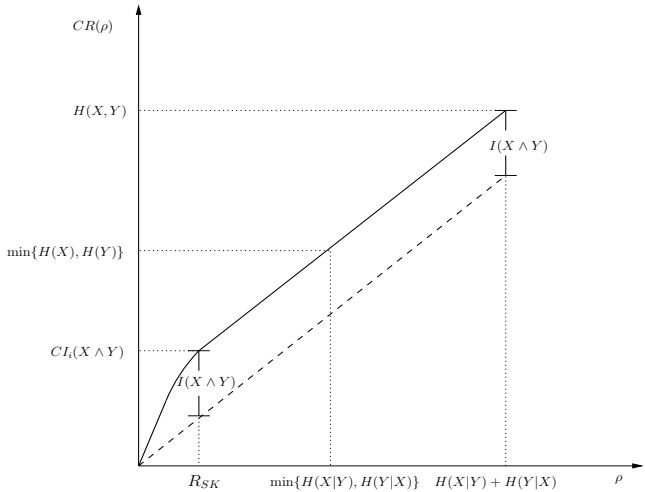, height = 6.5cm, width = 8.5cm}
\caption{Minimum rate of communication $R_{SK}$ for optimum rate
SK generation} \label{f_1}
\end{center}
\end{figure}

\noindent The graph of $CR$ as a function of  $\rho$ is plotted in Fig. \ref{f_1}. 
$CR(\rho)$ is an increasing and a concave function of $\rho$, as seen
from a simple time-sharing
argument. Since $R_{SK}$ is the minimum rate of communication 
required to generate a maximum rate SK, 
$CR(\rho) - \rho = I(X\wedge Y)$ for $\rho \geq R_{SK}$. 
Thus, our results characterize the graph of $CR(\rho)$  for
all $\rho \geq R_{SK}$. The quantity
$R_{SK}$ is the minimum value of $\rho$ for which the slope of 
$CR(\rho)$ is $1$; 
$CR\left(R_{SK}\right)$ is equal to the interactive common information
$CI_i(X\wedge Y)$. Furthermore,
from the proof of Theorem \ref{t_main},
a CR $L$ that satisfies (\ref{e_CIcond}) must 
yield an optimum rate SK. Thus, any CR recoverable from a
communication of rate less than $R_{SK}$ cannot satisfy 
(\ref{e_CIcond}). A characterization of $CR(\rho)$ for 
$\rho < R_{SK}$ is central to the characterization of $C(\rho)$,
and this, along with a single-letter characterization of $R_{SK}$,
remains an interesting open problem.


\section*{Appendix}
\setcounter{equation}{0}
\renewcommand{\theequation}{A\arabic{equation}}

\noindent{\it Proof of Lemma \ref{l:int_comm_rate}:}

From the Slepian-Wolf theorem \cite{SleWol73}, there exist mappings
$f_1, ..., f_r$ of $F_1^k, ..., F_r^k$, respectively, of rates
\begin{align}\nn
\frac{1}{k}\log \|f_{2i+1}\| &\leq H(F_{2i+1}\mid Y^n,
F_1,..., F_{2i}) + \frac{n\ep}{2r},
\\\nn &\hspace*{2.3cm} 0\leq i\leq \lfloor
(r-1)/2\rfloor,
\\\nn
\frac{1}{k}\log \|f_{2i}\| &\leq H(F_{2i}\mid X^n,
F_1,..., F_{2i-1}) + \frac{n\ep}{2r},
\\\nn &\hspace*{3.1cm} 1\leq i\leq \lfloor
r/2\rfloor,
\end{align}
such that
\begin{align}\nn
&F_{2i+1}^k \text{ is $\frac{\epsilon}{2r}$-recoverable
from } 
\\\nn &\hspace*{0.8cm} \left(f_{2i+1}(F_{2i+1}^k),Y^N, F_1^k,..., F_{2i}^k\right), 0\leq i\leq \lfloor (r-1)/2\rfloor,
\\\nn
&F_{2i}^k \text{ is $\frac{\epsilon}{2r}$-recoverable
from } 
\\\nn &\hspace*{0.8cm} \left(f_{2i}(F_{2i}^k),X^N, F_1^k,..., F_{2i-1}^k\right),
\quad 1\leq i\leq \lfloor r/2\rfloor,
\end{align}
for all $k$ sufficiently large. Thus, the communication 
$\mathbf{F}^\prime$ given by $F^\prime_i = f_i\left(F_i^k\right)$, 
$1\leq i \leq r$ constitutes the required communication of rate
\begin{align}\nn
\frac{1}{nk}\log\|\mathbf{F}^\prime\| \leq \frac{1}{n}\left[H\left(\mathbf{F}|X^n\right) + H\left(\mathbf{F}|Y^n\right)\right] + \ep.
\end{align}
\qed

\noindent{\it Proof of Lemma \ref{l:decomp}:}

For $T = T\left(X^n, Y^n\right)$ we have,
\begin{align}\nn
& nI(X\wedge Y) 
\\\nn &= H\left(X^n, Y^n\right)  - H\left(X^n\mid
Y^n\right) - H\left(Y^n\mid X^n\right)
\\\nn &=  H\left(X^n, Y^n\mid T\right)  - H\left(X^n\mid Y^n, T\right) - H\left(Y^n\mid X^n, T\right)
\\\nn &\qquad +H(T) - H\left(T\mid X^n\right)  - H\left(T\mid Y^n\right)
\\\nn &=  I\left(X^n\wedge Y^n\mid T\right) + H(T) - H\left(T\mid X^n\right)  - H\left(T\mid Y^n\right).
\end{align}
Lemma \ref{l:decomp} follows upon choosing $T = J, \mathbf{F}$.\qed

\vspace*{0.3cm}

\noindent{\it Proof of (\ref{el1}) and (\ref{el11}):}

It remains to prove that there exists $\ep$-CR $J$, recoverable
from $\mathbf{F}$ such that $J, \mathbf{F}$ satisfy (\ref{el1})
and (\ref{el11}). We provide a CR generation scheme with $r$
stages. For $1 \leq k \leq r$, denote by $\cE_{k}$ the error event 
in the $k$th stage (defined below recursively in terms of $\cE_{k-1}$), 
and by $\cE_0$ the negligible probability event corresponding to
$X^n, Y^n$ not being $P_{XY}$-typical. 

Consider $1 \leq k \leq r$, $k$ odd. For brevity, denote by
$V$ the rvs $U^{k-1}$ and by $U$  the rv $U_k$; for $k=1$,
$V$ is taken to be a constant. Suppose that conditioned on
$\cE_{k-1}^c$ terminals $\cX$ and $\cY$ observe, respectively,
sequences $\bx\in \cX^n$ and $\by\in \cY^n$, as well as a common
sequence $\bv \in \cV^n$ such that $(\bv, \bx, \by)$
are jointly $P_{VXY}$-typical. 
For $\delta>0$, generate at random 
$\exp\left[n(I(X, Y\wedge U\mid V) + \delta)\right]$ sequences
$\bu \in \cU^n$ that are jointly $P_{UV}$-typical with $\bv$, denoted
by $\bu_{ij}$, $1\leq i \leq N_1$, $1\leq j \leq N_2$, where 
\begin{align}\nn
N_1 &= \exp\left[n\left(I(X\wedge U\mid Y, V) + 3\delta\right)\right],
\\\nn
N_2 &= \exp\left[n\left(I(Y\wedge U\mid V) - 2\delta\right)\right].
\end{align}
The sequences $\bu_{ij}$ are generated independently for different 
indices $ij$. Denote by $L^{(k)}(\bv, \bx)$ a sequence $\bu_{ij}$,
$1\leq i \leq N_1$, $1\leq j \leq N_2$, that is jointly $P_{UVX}$-typical
with $(\bv, \bx)$ (if there exist more than one such sequences, choose
any of them). The error event when no such sequence is found is denoted
by $\cE_{k1}$; this happens with 
probability vanishing to $0$ doubly exponentially in $n$. 
The communication
$F_k(\bv, \bx)$ is defined to equal the first index $i$ of $\bu_{ij} = L^{(k)}(\bv, \bx)$.
Upon observing $F_k(\bv, \bx) = i$, the terminal $\cY$ computes 
$L_2^{(k)}(\bv, \by, i)$ as the unique sequence in $\{\bu_{ij}, 1\leq j \leq N_2\}$,
that is jointly typical with $(\bv, \by)$. If no such sequence is found or
if several such sequences are found an error event $\cE_{k2}$ occurs. 
Clearly, the rate of communication $F_k$ is bounded above by 
\begin{align}\nn
\frac{1}{n}\log N_1 &= I(X\wedge U\mid Y, V) + 3\delta
\\\label{ea2} &= I(X\wedge U_k\mid Y, U^{k-1}) + 3\delta,
\end{align}
and also, for large $n$,
\begin{align}\nn
\frac{1}{n}H(L^{(k)}) &\leq \frac{1}{n}\log(1 + N_1N_2) \leq I(X, Y \wedge U\mid V) + 2\delta
\\\label{ea3} &= I(X,Y \wedge U_k\mid Y, U^{k-1}) + 2\delta.
\end{align}
Denote by $\cE_{k3}$ the event $\left(L^{(k)}(\bv, \bx), \bv, \bx, \by\right)$
not being jointly $P_{UVXY}$-typical. The error event $\cE_k$ is defined as
$\cE_k = \cE_{k-1} \cup\cE_{k1}\cup \cE_{k2}\cup \cE_{k3}$.
Then, conditioned on $\cE_{k}^c$ the terminals 
share sequences $(\bu_{ij}, \bv)$ that are jointly typical with $(\bx, \by)$.
In the next stage $k+1$, the sequence $(\bu_{ij}, \bv)$ plays the role of the sequence $\bv$.
The scheme for stages with even $k$ is defined analogously with roles of $\cX$ and $\cY$ interchanged.
We claim that $L^{(1)}, ..., L^{(r)}$ constitutes the required CR along with the communication 
$\bF=F_1, ..., F_k$. Then, (\ref{el11}) follows from (\ref{ea2}), 
and the second inequality in (\ref{el1}) follows from (\ref{ea3}). Moreover,
for every realization $\bu_1, ..., \bu_r$ of $L^{(1)}, ..., L^{(r)}$, with $E = \mathbf{1}_{\cE_r}$ we have,
\begin{align}\nn
&\bPr{L^{(1)}, ..., L^{(r)} = \bu_1, ..., \bu_r \mid E = 0} 
\\\nn &\leq \bPr{\left\{(\bx, \by) : (\bu_1, ..., \bu_r, \bx, \by) \text{ are jointly $P_{U^r X Y}$ typical} \right\}}
\\\nn &\leq \exp\left[-n(I(X, Y \wedge U^r)  - \delta)\right],
\end{align}
for $n$ large, which further yields
\begin{align}\nn
\frac{1}{n}H(L^{(1)}...L^{(r)}\mid E = 0) \geq I(X, Y \wedge U^r) - \delta.
\end{align}
Therefore,
\begin{align}\nn
&\frac{1}{n}H(L^{(1)}...L^{(r)}) 
\\\nn &\geq \frac{1}{n}H(L^{(1)}...L^{(r)}\mid E = 0) - \bPr{\cE_r}\log|\cX||\cY|
\\\nn &\geq I(X, Y \wedge U^r) - \delta - \bPr{\cE_r}\log|\cX||\cY|.
\end{align}
Thus, the claim will follow upon showing that $\bPr{\cE_r} \rightarrow 0$ as $n \rightarrow \infty$.
In particular, it remains to show that $\bPr{\cE_{k2}} \rightarrow 0$ and $\bPr{\cE_{k3}} \rightarrow 0$,
$k = 1, ..., r$, as $n \rightarrow \infty$. As before, we show this for odd $k$ and the proof for even $k$
follows {\it mutatis mutandis}. To that end, note first that for any
jointly $P_{UVX}$-typical $(\bu, \bv, \bx)$, the set of $\by\in \cY^n$
such that $(\bu, \bv, \bx, \by)$ are jointly typical with
$(\bu, \bv, \bx)$ has conditional probability close to $1$
conditioned on $U^n= \bu, V^n= \bv, X^n =\bx$, and so by
the Markov relation $Y \mc V, X \mc U$, also conditioned on 
$V^n= \bv, X^n =\bx$. Upon choosing $\bu = L^{(k)}(\bv, \bx)$ in 
the argument above, we get $\bPr{\cE_{k2}} \rightarrow 0$. Finally, 
we show that $\bPr{\cE_{k3}}$ 
will be small, for large probability choices of
the random codebook $\{\bu_{ij}\}$. Specifically,
for fixed typical sequences $(\bv, \bx, \by)$, the 
probability
$\bPr{\cE_{k3}\mid V^n=\bv, X^n = \bx, Y^n = \by}$ 
is bounded above exactly as in \cite[equation (4.15)]{AhlCsi98}:
\begin{align}\nn
&\bPr{\cE_{k3} \mid  V^n=\bv, X^n = \bx, Y^n = \by} 
\\\nn &\leq \sum_{i=1}^{N_1}\sum_{j=1}^{N_2}\sum_{l=1, l\neq j}^{N_2}
\mathbb{P}\bigg((\bu_{ij}, \bv, \bx) 
\text{ jointly $P_{UVX}$-typical},
\\
\nn &\hspace*{3.5cm}
(\bu_{il}, \bv, \bu_{il}) \text{ jointly $P_{UVY}$-typical}\bigg)
\\\nn &\leq N_1N_2^2.\exp[-n(I(X\wedge U\mid V) + I(X\wedge U\mid V) + o(n))]
\\\nn &\leq \exp[-n\delta + o(n)],
\end{align}
for all $n$ sufficiently large. Note that the probability distribution in the calculation above
comes from codebook generation, and in particular, the second inequality above uses
the fact that $\bu_{il}$ and $\bu_{ij}$ are independently selected for 
$l \neq j$. Thus, 
$\bPr{\cE_{k3}\mid \cE_{k2}} \rightarrow 0$ for an appropriately chosen codebook,
which completes the proof. \qed

\section*{Acknowledgements}
The ideas presented in this work are based on 
heuristics for the interplay between CR generation and SK generation,
developed, over the years, jointly with Prof. Prakash Narayan.
Further, his comments on an earlier version of this manuscript
have improved the presentation, especially that of Section \ref{s_suff},
where he also simplified the proofs.


\end{document}